# Network Community Detection: A Review and Visual Survey

**Bisma S. Khan[1] · Muaz A. Niazi[*,1]**

[1]Department of Computer Science, COMSATS Institute of Information Technology, Islamabad, Pakistan
*Corresponding author. E-mail: muaz.niazi@gmail.com

**Abstract:** Community structure is an important area of research. It has received a considerable attention from the scientific community. Despite its importance, one of the key problems in locating information about community detection is the diverse spread of related articles across various disciplines. To the best of our knowledge, there is no current comprehensive review of recent literature which uses a scientometric analysis using complex networks analysis covering all relevant articles from the Web of Science (WoS). Here we present a visual survey of key literature using CiteSpace. The idea is to identify emerging trends besides using network techniques to examine the evolution of the domain. Towards that end, we identify the most influential, central, as well as active nodes using scientometric analyses. We examine authors, key articles, cited references, core subject categories, key journals, institutions, as well as countries. The exploration of the scientometric literature of the domain reveals that Yong Wang is a pivot node with the highest centrality. Additionally, we have observed that Mark Newman is the most highly cited author in the network. We have also identified that the journal, "Reviews of Modern Physics" has the strongest citation burst. In terms of cited documents, an article by Andrea Lancichinetti has the highest centrality score. We have also discovered that the origin of the key publications in this domain is from the United States. Whereas Scotland has the strongest and longest citation burst. Additionally, we have found that the categories of "Computer Science" and "Engineering" lead other categories based on frequency and centrality respectively.

**Keywords** Complex networks; community detection; CiteSpace; scientometric; visual survey

## 1. Introduction

Complex networks are the extremely important area of research. With the advancement in science and technology, a variety of research in the domain of complex networks has garnered a substantial amount of attention from the scientific community. Complex networks are expanding at a brisk pace. The growth of complex networks ranges from biological networks (Dunne et al. 2002; Jeong et al. 2000) to technological networks (Faloutsos et al. 1999; Albert et al. 1999; Amaral et al. 2000), from social networks (Wasserman and Faust 1994; Scott and Carrington 2011) to information networks (Newman 2004b). Complex networks are made up of interconnected nodes. With the increase in size and complexity of complex networks, it is essential to understand the related literature and key findings.

One of the key research fronts in this domain is community structures. Community structure is the most widely studied structural features of complex networks. Communities in a network are the dense groups of the vertices, which are tightly coupled to each other inside the group and loosely coupled to the rest of the vertices in the network. Community detection plays a key role in understanding the functionality of complex networks.

Recently, community detection has attracted a huge consideration due to the growing availability of the data sets of the large-scale networks. To provide insightful information about community detection, much research has been conducted in the form of surveys, systematic literature reviews, and visual studies. But, only a few of them shows how the field advanced over time. To demonstrate the sense of details, information about existing literature is listed in Table 1.

**Table 1.** The existing literature review in the domain of "Network Community Detection"

| **Ref.** | **Paper Type** | **Study Area** |
|---|---|---|
| (Cai et al. 2016) | Survey | Evolutionary techniques for the identification communities in networks |
| (Fortunato and Hric 2016) | User Guide | Identification of communities in networks |
| (Bedi and Sharma 2016) | Advanced Review | Identifying communities in social networks |
| (Enugala et al. 2015) | Survey | Uncovering communities in dynamic social networks |
| (Dhumal and Kamde 2015) | Survey | Community discovery in online social networks |
| (Drif and Boukerram 2014) | Literature Survey | Dynamic community identification and social network models |
| (Dhumal and Kamde 2015) | Survey and empirical evaluation | Community identification in large-scale networks |
| (Drif and Boukerram 2014) | Survey | Techniques for uncovering communities in social networks |
| (Y.-X. Ma et al. 2013) | Visual Analysis | Community discovery of multi-context mobile social networks |
| (Plantié and Crampes 2013) | Survey | Social community identification |
| (Malliaros and Vazirgiannis 2013) | Survey | Community discovery in directed networks |
| (Coscia et al. 2011) | Review | Classification for community detection approaches in social networks |
| (Fortunato 2010) | Survey | Identification of communities in graphs |

In spite of these studies, one of the key difficulty researchers face in locating information about community detection is that numerous important references are found in different related disciplines – primarily due to the multidisciplinary landscape associated with the domain. Therefore, it is quite difficult to get acquainted with the basic concepts, historical trends, general developments in the field, and future directions.

In this paper, we explore bibliographic literature from Web of Science (WoS) using CiteSpace, to get insight into publication performance of research in the domain, to trace its temporal evolution, and to identify its intellectual structure using visual analysis. Formerly, CiteSpace has been used in various fields, such as agent-based computing (Niazi and Hussain 2011), visualisation of aggregation operator (Yu 2015), anticancer research (Xie 2015), digital divide (Zhu et al. 2015), digital medicine (Fang 2015), and tech mining (Madani 2015), etc. Additionally, we have used Pajek for structural analysis. To the best of our knowledge, until now, there is no current review of recent literature on community detection, which uses a scientometric analysis of networks formed from highly cited and important journal papers from the Web of Science (WoS) to investigate the general development of the domain.

The key contribution of this paper is the identification of the emerging trend, structure, and evolution in the domain of community detection through exploring central nodes, landmark nodes, and bursting nodes from scientometric literature. The ideas of visual analysis and survey stem from Cognitive Agent-based Computing framework (Niazi and Hussain 2011) – a framework which allows for modelling and analysis of natural and artificial Complex Adaptive Systems.

The key results of our study are as follows. First, we have explored most productive authors based on visual analysis of the author collaborative and co-cited networks. Then, we have revealed key articles through co-citation analysis of documents, journals, institutions, and countries of the origin of the manuscripts. Next, we seek core subject categories of the domain. Subsequently, we identified structural patterns and developments in the domain through the exploration of clusters of coauthors, cited documents, and cited journals.

The rest of the paper is structured as: Section II explores the background of network community detection techniques. Section III describes the adopted methodology. Section IV presents the results and discussion. Section VI draws conclusions.

## 2. Background

This section presents the necessary background of network community detection for better understanding.

### 2.1. Graph Theoretical Notation of the Network

A complex network can be mapped to the graph $G(V, E)$, where $V$ is the node set and $E$ is the edge set. The cardinality (order) of $V$ and $E$ is represented by $n$ and $m$ respectively. A network $C(v, e)$ is said to be subnetwork if $v$ is the subset of $V$ and $e$ is the subset of $E$.

Let A is the adjacency matrix; two nodes are adjacent if they have a link between them. If there exists a link between vertex i and j then $A_{ij} = 1$, otherwise $A_{ij} = 0$. A weighted network has weight $w$ attached to the edges, where $w$ is a real number.

Let $K_i = \sum_{j=1}^{N} A_{ij}$ denotes the degree of a vertex i that is a total number of links incident to vertex i. Directed networks contain two types of degrees: a) in-degree and b) out-degree. The number of arcs a node receives is called in-degree, such that $k_i^{in} = \sum_{j=1} A_{ij}$. Whereas the number of arcs a vertex sends is called out-degree, such that $k_i^{out} = \sum_{j=1} A_{ji}$. Total degree in directed networks is $K_i = k_i^{in} + k_i^{out}$, whereas in undirected networks total degree is $K_i = k_i^{in} = k_i^{out}$. In an undirected network, the total number of degrees is double the number of links in the network.

"Network density is the number of lines in a simple network, expressed as a proportion of the maximum possible number of lines" (De Nooy et al. 2011).

The internal degree $k_i^{int}$ of vertex i belongs to $C$, is the number of links connecting vertex i to the other members of $C$, such that $k_i^{int} = \sum_{j \in C} A_{ij}$. The external degree of vertex i is the number of edges connecting vertex i to rest of the nodes in $G$ outside $C$, such that $k_i^{ext} = \sum_{j \notin C} A_{ij}$.

## 2.2. Community Detection

Modern networks are growing exponentially in size, variety, and complexity. As a result of changes in the networks, newer and different types of communication networks are emerging, such as multi-agent, Internet of Things, ad-hoc, wireless sensors, cloud-based, co-citation, and social networks. Network operations inherent several nonlinearities, which leads to the rise in complex emergent patterns. These patterns are important to understand because they can have unexpected effects on various characteristics of the network.

Communities in a network are the groups of nodes, which are highly connected to each other than to the rest of the nodes in the network (Yang et al. 2010). Community detection is the key characteristic, which could be used to extract useful information from networks. The greatest challenge in community detection is that no universal definition of community structure exists (Fortunato and Hric 2016). Therefore, community detection in large-scale networks is computationally intractable.

A large number of techniques has been suggested to find optimal communities in reasonably fast time. Most of these techniques are based on the optimisation of objective functions. Modularity optimisation by so far is one of the most widely used techniques among them. However, modularity optimisation is an NP-hard problem.

### 2.2.1. Modularity Q

This section briefly describes modularity, for further details readers are encouraged to see (Fortunato 2010). Modularity $Q$ is the measure of the density of intra-community links as compared to inter-community links.

The original idea of modularity was given by Newman and Girvan (Newman 2004a), they have defined modularity Q as:

$$Q = \frac{1}{2m} \sum_{i,j} \left[ A_{ij} - \frac{k_i k_j}{2m} \right] \delta(C_i, C_j) \qquad (1)$$

Here, $m$ is the number of links, $k_i$ is the degree of vertex i, $k_j$ is the degree of vertex j, $C_i$ is the community to vertex i, $C_j$ is the community to vertex j, and $\delta(C_i, C_j) = 1$ if i and j belong to the same community, otherwise it equals to $0$.

Newman and Girvan (Newman and Girvan 2004) have also defined modularity Q as:

$$Q = \sum_{k=1}^{s} \left[ \frac{l_k}{L} - \left( \frac{d_k}{2L} \right)^2 \right] \qquad (2)$$

Here, $s$ is the number of modules, L is the sum of all links in the network, $l_k$ is the number of internal edges of a community $k$, and $d_k$ is the sum of the degrees of all vertices in the community $k$.

In the case of weighted networks, Newman (Newman 2004a) has defined modularity Q as:

$$Q = \frac{1}{2W}\sum_{i,j}\left[A_{ij} - \frac{s_i s_j}{2W}\right]\delta(C_i, C_j) \quad (3)$$

Here, W is total weight of all of the links in the network, $W_{ij}$ is the weight of the links between vertices i and j, $s_i$ is the strength of vertex i, $s_j$ is the strength of vertex j, and $\delta(C_i, C_j) = 1$ if i and j belong to the same community, otherwise it equals to 0.

Which can also be defined as:

$$Q = \sum_{c=1}^{n_c}\left[\frac{W_c}{W} - \left(\frac{S_c}{2W}\right)^2\right] \quad (4)$$

Here, $S_c$ is the total strength of the nodes of community c and $W_c$ is the total weight of the internal edges of c.

In the case of directed networks, Arenas et al. (Arenas et al. 2007) have defined modularity Q as:

$$Q = \frac{1}{2m}\sum_{i,j}\left[A_{ij} - \frac{k_i^{out}k_j^{in}}{m}\right]\delta(C_i, C_j) \quad (5)$$

Here $A_{ij} = 1$ if link between nodes i and j, $k_i^{out}$ is the out-degree of vertex i, $k_j^{in}$ is the in-degree of vertex j, $C_i$ is the community membership of vertex i, $C_j$ is the community membership of vertex j, and $\delta(C_i, C_j) = 1$ if i and j belong to the same community, otherwise it equals 0.

In the case of weighted and directed networks, Arenas et al. (Arenas et al. 2007) have defined modularity Q as:

$$Q = \frac{1}{2W}\sum_{i,j}\left[W_{ij} - \frac{k_i^{out}k_j^{in}}{W}\right]\delta(C_i, C_j) \quad \text{(Erdos and Rényi)}$$

For overlapping communities in the case of unweighted and undirected networks Shen et al. (Shen et al. 2009) have defined modularity Q as:

$$Q = \frac{1}{2m}\sum_{i\,j}\frac{1}{O_i O_j}\left(A_{ij} - \frac{k_i k_j}{2m}\right)\delta(C_i, C_j) \quad (7)$$

Here $O_i$ is the number of modules containing node i, $O_j$ is the number of modules including node j.

For overlapping communities in the case of unweighted and directed networks, Nicosia et al. (Nicosia et al. 2009) have defined modularity Q as:

$$Q = \frac{1}{m}\sum_{c=1}^{n_c}\sum_{i,j}\left[r_{ijc}A_{ij} - s_{ijc}\frac{k_i^{out}k_j^{in}}{m}\right] \quad (8)$$

| Ref. | Modularity | Network Type | | Community Type |
|---|---|---|---|---|
| | | Directed | Weighted | |
| (Newman 2004a) | $Q = \frac{1}{2m}\sum_{i,j}\left[A_{ij} - \frac{k_i k_j}{2m}\right]\delta(C_i, C_j)$ | No | No | Disjoint |
| (Newman and Girvan 2004) | $Q = \sum_{k=1}^{s}\left[\frac{l_k}{L} - \left(\frac{d_k}{2L}\right)^2\right]$ | No | No | Disjoint |
| (Newman 2004a) | $Q = \frac{1}{2W}\sum_{i,j}\left[A_{ij} - \frac{s_i s_j}{2W}\right]\delta(C_i, C_j)$ | No | Yes | Disjoint |
| (Fortunato 2010) | $Q = \sum_{c=1}^{n_c}\left[\frac{W_c}{W} - \left(\frac{S_c}{2W}\right)^2\right]$ | No | Yes | Disjoint |
| (Arenas et al. 2007) | $Q = \frac{1}{2m}\sum_{i,j}\left[A_{ij} - \frac{k_i^{out}k_j^{in}}{m}\right]\delta(C_i, C_j)$ | Yes | No | Disjoint |
| (Arenas et al. 2007) | $Q = \frac{1}{2W}\sum_{i,j}\left[W_{ij} - \frac{k_i^{out}k_j^{in}}{W}\right]\delta(C_i, C_j)$ | Yes | Yes | Disjoint |
| (Shen et al. 2009) | $Q = \frac{1}{2m}\sum_{ij}\frac{1}{O_iO_j}\left(A_{ij} - \frac{k_ik_j}{2m}\right)\delta(C_i, C_j)$ | No | No | Overlapping |
| (Nicosia et al. 2009) | $Q = \frac{1}{m}\sum_{c=1}^{n_c}\sum_{i,j}\left[r_{ijc}A_{ij} - s_{ijc}\frac{k_i^{out}k_j^{in}}{m}\right]$ | Yes | No | Overlapping |

Here $c$ is the indexing label of communities and $r_{ijc}$, $s_{ijc}$ represent the contributions to the sum corresponding to the link ij in the network and in the null model, because of the multiple memberships of i and j. Table 2 lists various variations of modularity for different types of networks.

**Table 2.** Different variations of modularity for different types of networks

### 2.2.2. Modularity Density

To resolve community structure of networks, Li et al. (Zhenping Li et al. 2008) have introduced a new quantitative measure named as modularity density (D), which is based on the density of subgraphs. The higher the value of D, the better is a partition. The optimisation of modularity density is also NP-hard. The modularity density is defined as:

$$D_\lambda = \sum_{i=1}^{m}\frac{2\lambda L(V_i,V_i) - 2(1-\lambda)L(V_i-\overline{V_i})}{|V_i|} \quad (9)$$

Where, $V_i$ is the subset of $V$ $i = 1, \ldots, m$, such that $L(V_i, V_i) = \sum_{i\in V_i, j\in V_i} A_{ij}$ and $L(V_i, \overline{V_i}) = \sum_{i\in V_i, j\in \overline{V_i}} A_{ij}$, where $\overline{V_i} = V - V_i$

### 2.2.3. Most commonly used real-networks/Datasets

Table 3 shows most commonly used data set for identification of communities.

**Table 3.** Most commonly used dataset in community detection algorithms

| Dataset | Reference | No. of Nodes | No. of Links |
|---|---|---|---|
| Zachary karate club | (Zachary 1977) | 34 | 78 |
| American college football network | (Girvan and Newman 2002) | 115 | 616 |
| Southern women dataset | (Davis et al. 1941) | 18 | - |
| Lusseau's dolphins' network | (Lusseau et al. 2003) | 62 | 159 |
| JAZZ musician network | (Gleiser and Danon 2003) | 198 | 2,742 |
| C. metabolic network | (Jeong et al. 2000) | 453 | - |
| Condense Matter collaboration network (Cond-Mat) | (Newman 2001a) | 23133 | 9,3497 |
| Actor Movie | (A.-L. Barabási and Albert 1999) | 233,283 | 2,81,396 |
| Corporate interlocks in Scotland | (Scott and Hughes 1980) | 244 | 356 |
| ego-Facebook | (McAuley and Leskovec 2012) | 4,039 | 88,234 |
| LiveJournal | (Backstrom et al. 2006) | 4000000 | 3,4900000 |
| Lancichinetti–Fortunato–Radicchi (Cuzzocrea et al.) | (Lancichinetti et al. 2008) | - | - |
| University E-mail network | (Guimera et al. 2003) | 1169 | - |
| PGP | (Guardiola et al. 2002) | 10680 | - |
| Pollbooks | (Krebs 2004) | 105 | - |
| Amazon | (Leskovec et al. 2007) | 8275 | 22,231 |
| DBLP | (Backstrom et al. 2006) | 26,956 | 88,742 |
| Orkut | (Mislove et al. 2007) | 297,691 | 7,747,026 |
| Collaboration network | (Girvan and Newman 2002) | 271 | - |

### 2.3. Community Detection Techniques

Identification of community structure can help us to understand network functionality. However, community detection is computationally intractable in large-scale networks. Despite the substantial interest of scientific community on it over last few years, no universally accepted solution of community detection exists yet. Thus far, a considerable number of techniques for the optimisation of community detection has been introduced in the scientific literature.

In this section, we present an overview of the six state-of-the-art groups of community detection algorithms. We provided a brief overview of techniques for the identification of static, dynamic, disjoint, and overlapping communities. To the best of our knowledge, the list comprises of the most current community detection algorithms. The comprehensive overview of the community detection techniques can be found in (Cai et al. 2016; Fortunato and Hric 2016; Fortunato 2010).

#### 2.3.1. Traditional Community Detection Techniques

Here we highlight traditional techniques of community detection.

##### 2.3.1.1. Graph partitioning

It divides the graph into g clusters of predefined size, such that the number of links in a cluster is more denser than the number of edges between the clusters (Fortunato 2010). Well-known examples of graph partitioning techniques are Spectral Bisection method (Barnes 1982) and Kernighan-Lin algorithm (Kernighan and Lin 1970).

##### 2.3.1.2. Hierarchical clustering

The graphs may contain hierarchical structure, that is each community may be a collection of small clusters at different levels (Fortunato 2010; Friedman et al. 2001). In such cases, hierarchical clustering techniques may be used to identify the multilevel community structure of the graph. Hierarchical clustering techniques are based on the vertex similarity measure. They do not need a predefined size and number of communities. They can be better represented by dendrograms. Hierarchical clustering techniques can be categorised into two classes:

- Agglomerative algorithms

  It is a bottom-up technique because at the start it considers each node as a separate cluster and iteratively merge them based on high similarity and ends up with the unique community.
- Divisive algorithms

  It is a top-down technique because at the start it considers the entire network as a single cluster and iteratively splits it by eliminating links joining nodes with low similarity and ends up with unique communities.

##### 2.3.1.3. Partitional clustering

Partitional clustering (Jin and Han 2011; Fortunato 2010; Dhumal and Kamde 2015; Furht 2010; Slaninová et al. 2010) partitions a dataset into a predefined number of k non-overlapping clusters. The goal is to divide the data points into k clusters in order to minimise/maximise the cost function based on dissimilarity measure between nodes. Some of the commonly used cost functions are minimum k-median, k-clustering sum, k-clustering, and k-center. Examples of partional clustering techniques include k-mean clustering (MacQueen 1967) and fuzzy k-mean (Bezdek 2013; Dunn 1973) clustering. Note that in fuzzy k-mean clustering one node may belong to multiple clusters.

##### 2.3.1.4. Spectral clustering

Spectral clustering includes all techniques which use eigenvectors of matrices to divide the set of data points based on the pairwise similarity between them (Fortunato 2010; Dhumal and Kamde 2015).

Examples include a Laplacian spectral partitioning method of Fiedler (Fiedler 1973) and Donath (Donath and Hoffman 1973).

#### 2.3.1.5. Divisive algorithms

It removes inter-cluster edges in a network based on low-similarity to separate communities from each other (Murata 2010). The main examples of this type include Girvan-Newman algorithm (Girvan and Newman 2002) where edges are removed iteratively based on edge-betweenness score and Radicchi et al. technique (Madani 2015), where edges are removed iteratively based on the edge clustering coefficient.

### 2.3.2. Modularity Optimisation Based Community Detection Techniques

This subsection presents network community detection techniques based on modularity optimisation. Modularity is the quality function for the approximation of communities. The larger the modularity value the better is the partition.

#### 2.3.2.1. Greedy techniques

i. Greedy method of Newman

Newman's greedy search algorithm (Newman 2004c) was the first algorithm suggested for modularity optimisation. It is an agglomerative technique, where initially, each node belongs to a distinct module, then they are merged iteratively based on the modularity gain. It has a time complexity of $O(n)^3$ on sparse networks.

ii. Fast Greedy algorithm (CNM)

It is the fast version of Newman's algorithm (Newman 2004c), implemented by efficient data structures. It has a computational complexity of $O(n\log^2 n)$ on sparse networks.

iii. Blondal's Louvain algorithm

Louvain (Clauset et al. 2004) is a heuristic greedy algorithm for uncovering communities in complex weighted graphs. It is also based on the modularity optimisation. It assigns different communities to each vertex; one per vertex. It iteratively merges the nodes based on the gain of modularity. If no gain, then node remains in its own community. The procedure is repeated until no more improvement is possible. It then reconstructs the network in the way that communities identified in the first phase are replaced by supernodes. Its time complexity is O(nlogn).

#### 2.3.2.2. Simulated Annealing (Kirkpatrick et al. 1983)

It is a discrete stochastic approach used for the global optimisation of the given objective function. Guimer`a et al. (Guimera and Amaral 2005) have used simulated annealing based modularity optimisation approach. Initially, it decomposes the network into random partitions. The optimisation is based on local and global moves. Local moves shift a node randomly from one partition to another based on modularity gain. Global moves consist of splitting and merging of partitions.

#### 2.3.2.3. Extremal Optimisation

Boettcher et al. (Boettcher and Percus 2001) have designed extremal optimisation as a general purpose heuristic search technique for physical and combinatorial optimisation problems. It is proposed to gain an accuracy comparable with genetic algorithm and simulated annealing. It focuses on the optimisation of local variables. Duch et al. in (Duch and Arenas 2005) have used it for modularity optimisation. It assigns fitness to each node; the fitness value is obtained by taking the ratio of the local modularity of the node by its degree. It evolves an individual solution within a single configuration and makes local modifications. It starts by randomly dividing the network into two partitions of the same order. It iteratively shifts vertices with the lowest fitness to the other partitions. After shifting, the partitions change, therefore it recalculates local fitness of many nodes. The process repeats until an optimum value of global modularity is reached.

#### 2.3.2.4. Spectral Optimisation

Spectral optimisation refers to the use of eigenvectors and eigenvalues of the modularity matrix for modularity optimisation (Fortunato 2010). This optimisation is quite fast. Examples of the spectral optimisation of modularity include (Richardson et al. 2009; Newman 2006).

#### 2.3.2.5. Evolutionary algorithms

Evolutionary algorithms are a class of metaheuristic optimisation algorithms based on artificial intelligence. They are known for their effective local learning and global searching capabilities. These methods are broadly divided into two classes based on single and multi-objective optimisation. Examples of the first category include MAGA-Net (Zhangtao Li and Liu 2016), MLAMA-Net (Mirsaleh and Meybodi 2016), MLCD (L. Ma et al. 2014), etc. Examples of the second category include MOEA/D (Zhang and Li 2007), COMBO (Sobolevsky et al. 2014), I-NSGAII (Deng et al. 2015), etc.

### 2.3.3. Overlapping community detection techniques

In real networks, most of the nodes may simultaneously belong to multiple communities. Traditional community detection techniques fail in identifying overlapping communities.

Clique percolation is the most known technique used for the identification of overlapping communities in the networks. It is based on the idea that cliques are more likely to be formed from internal edges which are densely connected than from external edges which are sparsely connected. The communities are made up of k-cliques which refer to the complete subgraphs with k vertices. Two cliques are known adjacent if they share $k-1$ nodes. The k-clique community is the giant component formed of all the adjacent k-cliques which are connected as a k-clique series. Other examples of this category include top graph clusters (Macropol and Singh 2010), SVINET (Gopalan and Blei 2013), and label propagation algorithm (Raghavan et al. 2007).

### 2.3.4. Dynamic Community Detection Algorithms

This subsection focuses on the community detection techniques in dynamic networks, such as Twitter, Facebook, LinkedIn, etc. These techniques revise the community assignment of the changed or new vertices during temporal updates in the network (Shang et al. 2016).

#### 2.3.4.1. Potts model

The Potts model is one of the well-known models used in statistical physics (Wu 1982). It demonstrates a system of spins which can be in q different states. The interaction among neighbouring spins may be ferromagnetic or antiferromagnetic. Potts spin variables can be mapped to the nodes of the graph having community structure. From interactions between neighbouring spins, it is plausible that community structure may be identified from like-valued spin clusters of the system, as there will be more interactions in the community and fewer interactions outside the community. Inspired by this idea, Reichardt et al. (Reichardt and Bornholdt 2004) have suggested a community detection technique which is based on the q-Potts model with interactions among nearest neighbours.

#### 2.3.4.2. Random walk

The Random Walk (Hughes 1996) is adopted to identify communities. In a random walk, the walker starts to walk inside a community from a node and at each time step it moves to the neighbouring node selected randomly and uniformly. The walker devotes a long time inside the dense communities because of high density and multiple paths. Examples of the most popular techniques based on random walks are PageRank algorithm (Page et al. 1999), WalkTrap (Pons and Latapy 2005), and Infomap (Rosvall and Bergstrom 2008).

#### 2.3.4.3. Diffusion Community

"A diffusion community in a complex network is a set of nodes that are grouped together by the propagation of the same property, action or information in the network (Coscia et al. 2011)." Examples of this category include label propagation (Raghavan et al. 2007) and dynamic node colouring (Tantipathananandh et al. 2007).

2.3.4.4. Synchronization

Synchronisation is an emerging phenomenon which has received interest from different fields. It occurs in interacting units and is persuasive in nature, technology and society. In a synchronised state, the system units remain in same or alike states over time. Synchronisation is also used in community detection in networks. If oscillators are located at nodes with random initial phases and interact with nearest neighbours, then oscillators belonging to the same partition synchronise first, whereas longer time is required for full synchronisation. Thus, if one follows the evolution time of the process, then states with synchronised communities of nodes can be much stable and prolonged, therefore can be identified easily. Examples of this type include (Arenas et al. 2006; Boccaletti et al. 2007).

## 3. Methodology

This section demonstrates how we collect data and visualise emerging trends and investigate landmark articles, influential authors and journals, central countries and institutions, and leading categories. It also explains how we identify largest clusters in the collaborative author's network, document co-citation networks, and journal citation network.

Figure 1 demonstrates research procedure for the visual scientometric analysis of the scientific literature in the domain of "network community detection" to identify dynamics of the domain. We have collected data from WoS using appropriate query. Then we have performed following analysis: firstly, we have analysed collaborative authors network and identified the largest cluster of co-authors. We have also investigated structural and bibliometric properties of the largest connected components. Next, we performed co-citation analysis of authors, journals, and documents. We have identified largest clusters of the cited authors, journals, and documents to explore underlying research themes. Next, we have investigated collaboration networks at institution and country level. Subsequently, we have performed co-occurrence analysis of categories to identify hot areas of research.

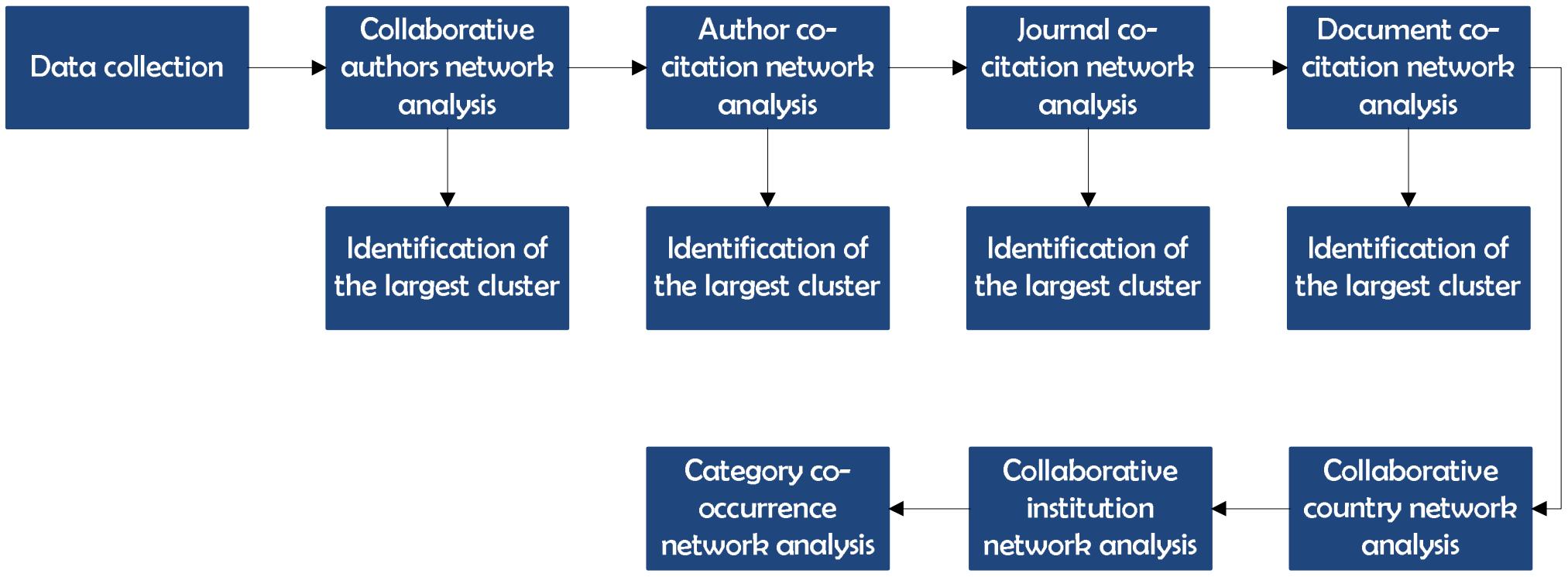

**Figure 1.** Research methodology (adapted from (M. C. Kim et al. 2016)) for the visual analysis of "network community detection" for the identification of emergent patterns and trends in the scientific data of the domain.

### 3.1. Data Collection

To perform significant co-citation analysis of bibliographic literature, we have collected input data from Thomson Reuters' Web of Science (WoS) on Nov 11, 2016. An extensive topic search was performed using ((communit*) AND (detect*) AND (network*)) as a query in the time frame of all years. A full record and cited references were selected as record content. The search was refined by selecting the document type as journal articles and language as English, which resulted in 3,168 unique records. For all years, a search was made in all four databases of WoS, including SSCI, ESCI, A&HCI, and SCI-EXPANDED. Our dataset contains 94.802% articles, 4.517% reviews, 4.455% proceedings papers, 0.495% editorial material, 0.155% book chapters, 0.093% letters, and 0.062% notes.

We start by examining the citation report of records retrieved from WoS with "communit* detect* network" as a keyword. Figure 2 depicts the growth rate of items published in latest 20 years. It can be observed that publications in this domain have increased from 15 publications in the year 1998 and reached an aggregate of 485 publications merely in the year 2015. Following it closely are 480 publications in the year 2016, which is a clear indicator of the interest of the research community in the domain of "network community detection" in recent years.

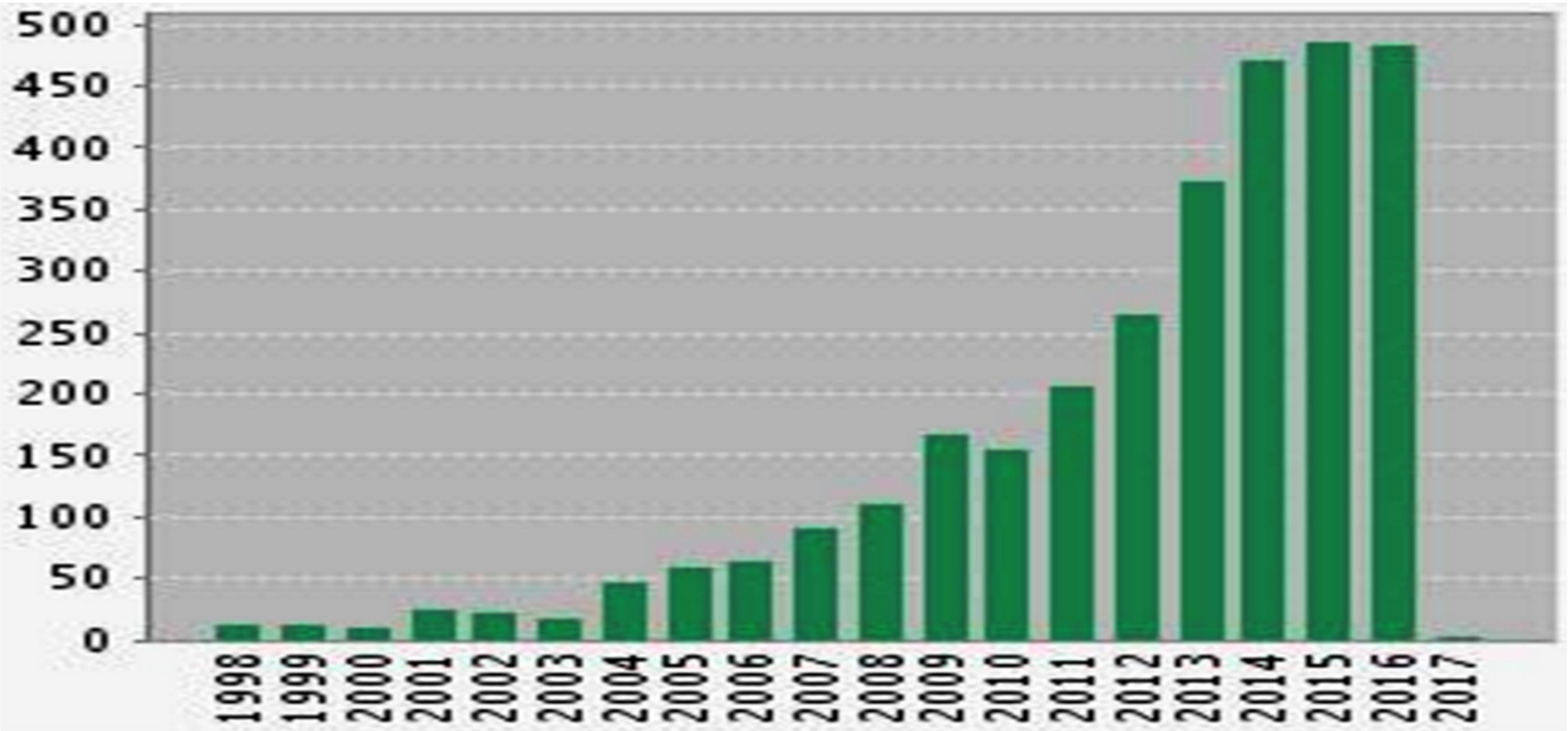

**Figure 2.** The chronological distribution of the number of articles in "network community detection" literature published over the past 20 years.

It is known that the popularity of a domain is directly proportional to the number of citations in that domain (Ding and Cronin 2011; Franceschet 2010). It can be viewed in Figure 3 that starting from the citations of 100 in 1998, the "network community detection" has risen to 13,182 citations alone in the year 2015 and approximately 10,900 citations in the year 2016. It clearly specifies that network community detection has gained much attention in current years.

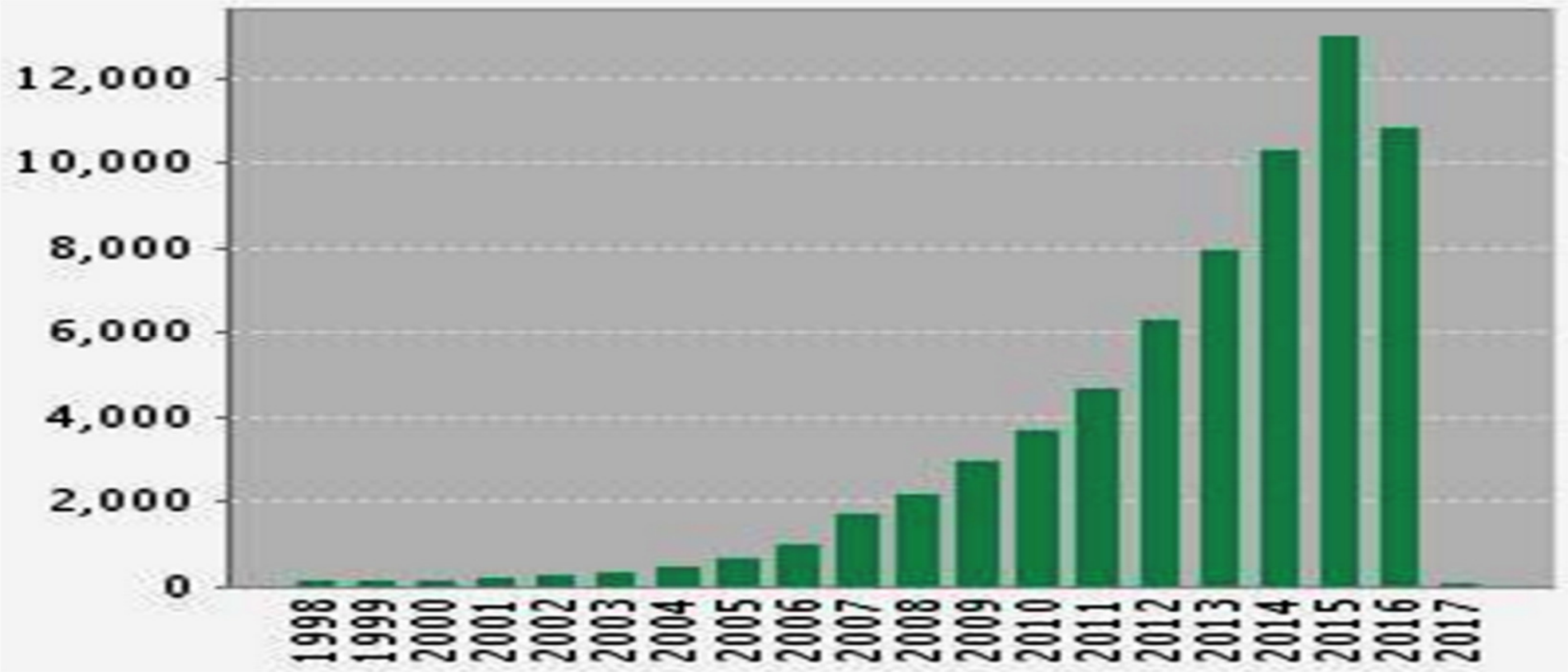

**Figure 3.** Citations history of "network community detection" bibliographic data over the latest twenty years.

The entire input dataset was then imported in CiteSpace for further analysis.

### 3.2. Network Analysis and Visualisation Tools

Several network visualisation and analysis tools are available, such as Pajek, Gephi, WoS2Pajek, etc. Pajek and Gephi are the most popular tools for general network analysis. However, they lack specific functionality for processing scientometric data. They require other software tools to extract data from WoS bibliographic database, such as Pajek uses WoS2Pajek and Gephi uses Sci2. Visualisation capabilities of Pajek are relatively weak whereas, Gephi has relatively strong visualisation capabilities. Another tool, UCInet uses Netdraw and Pajek for visualisation and is not freeware. VOSview is a tool developed specifically for the visual analysis of bibliometric networks. Unlike other tools, it offers distance-based visualisation of scientometric networks. It allows visualisation of larger networks;

however, it is relatively weak on analytics. It has memory constraints and computational limitations.

### 3.3. CiteSpace: an Overview

In this research, we have used CiteSpace, a Java-based key visual analytical tool for network analysis and visualisation (Chen 2006). It is tailored for the interactive visualisation of citation information obtained from the scientific literature. CiteSpace is captivated in clarification of the structure and dynamics of the domain. CiteSpace directly takes data downloaded from Web of Science and generates various panoramic networks. Based on the user's choice, a network can be viewed in terms of betweenness centrality, citation frequency, or citation burstness.

- "The *betweenness centrality* of a vertex is the proportion of all geodesics (shortest path) between pairs of other vertices that include this vertex" (De Nooy et al. 2011). In CiteSpace, betweenness centrality score ranges between $0$ and $1$. Nodes with high betweenness centrality are highlighted with purple trims. The thickness of the purple trim is proportional to the centrality score. The pink circle around the node represents the betweenness centrality score $>= 0.1$.
- "The *burstness* of the frequency of an entity over time indicates a specific duration in which an abrupt change of the frequency takes place (Chen 2014; Kleinberg 2003)". The red circle around the node represents the significant *citation burst*, indicating that citations of this node have emerged rapidly in a particular time period.

The node size is proportional to the overall frequency of citations. The concentric rings around the node indicate the temporal citation history of the publications. The colour of the citation ring indicates the citation in a particular time slice. A link between a pair of vertices in the network represents interaction. The thickness of a line indicates the co-authorship strength.

As CiteSpace is a visualisation tool, it extensively depends on colours, therefore the description in this paper is based on colours. The colour of a link indicates the time slice when the link was first established. Blue colour represents the earliest years, green colour represents the middle years, and orange and red colours represent the current years. The darker shades of the same colour indicate the earlier time-slice, whereas lighter shade indicates the later time slice.

The CiteSpace is custom designed to identify significant articles from the citations network. Articles identified this way are of certain importance, they may direct the future developments.

The four most generally used nodes for the identification of potentially important articles are hub nodes, landmark nodes, pivot nodes, and turning points.

- *Landmark nodes* are the highly-cited nodes with largest radii.
- *Hub nodes* are widely co-cited nodes with a relatively large degree.
- *Pivot nodes* act as exclusive joints between different clusters.
- *Turning points* are the highly central nodes standing between different groups of articles. "Turning points refer to the revolutionary articles identified by domain experts, whereas pivotal points refer to articles that share similar topological properties in the network generated by CiteSpace" (Chen 2006).

Although CiteSpace provides powerful visualisation, it is relatively weak in analytics. Therefore, we have used Pajek in addition to CiteSpace for structural analysis of the bibliographic networks.

## 4. Results and Discussion

### 4.1. Author Collaborative Network Analysis

We carried a visual analysis author collaborative network to find the core authors contributing to the literature of "network community detection." It is a network of co-authors, i.e. they have published articles together. Two authors are considered connected if they have co-authored a publication (Newman 2001b). In author collaborative network, authors are the nodes and co-authorship of papers are the links. The merged network as shown in Figure 4 contains $1414$ nodes and $2534$ links. We have selected top $30$ authors per one-year slice with an interval $[0, 5]$ of the citation distribution.

An edge between two authors in the network represents a collaboration. The thickness of a line indicates the co-authorship strength. The colour of a link indicates the time slice when the link was first established. The red circle around the node represents the significant citation burst, indicating that citations of this author have emerged rapidly in a particular time period. The node size is proportional to the overall frequency of citations. The concentric rings around the node indicate the temporal citation history of the publications. The colour of the citation ring indicates the citation in a particular time slice.

Figure 4 depicts that Liu J with largest radii is the most important researcher in the field of "network community detection." The red concentric circles around Jiao LC is the indicator of strongest citation burst. It provides the evidence that publications of Jiao LC have suddenly attracted a higher degree of attention from its research community. To spot salient features of the collaborative author network, detailed information is given below in tabular form.

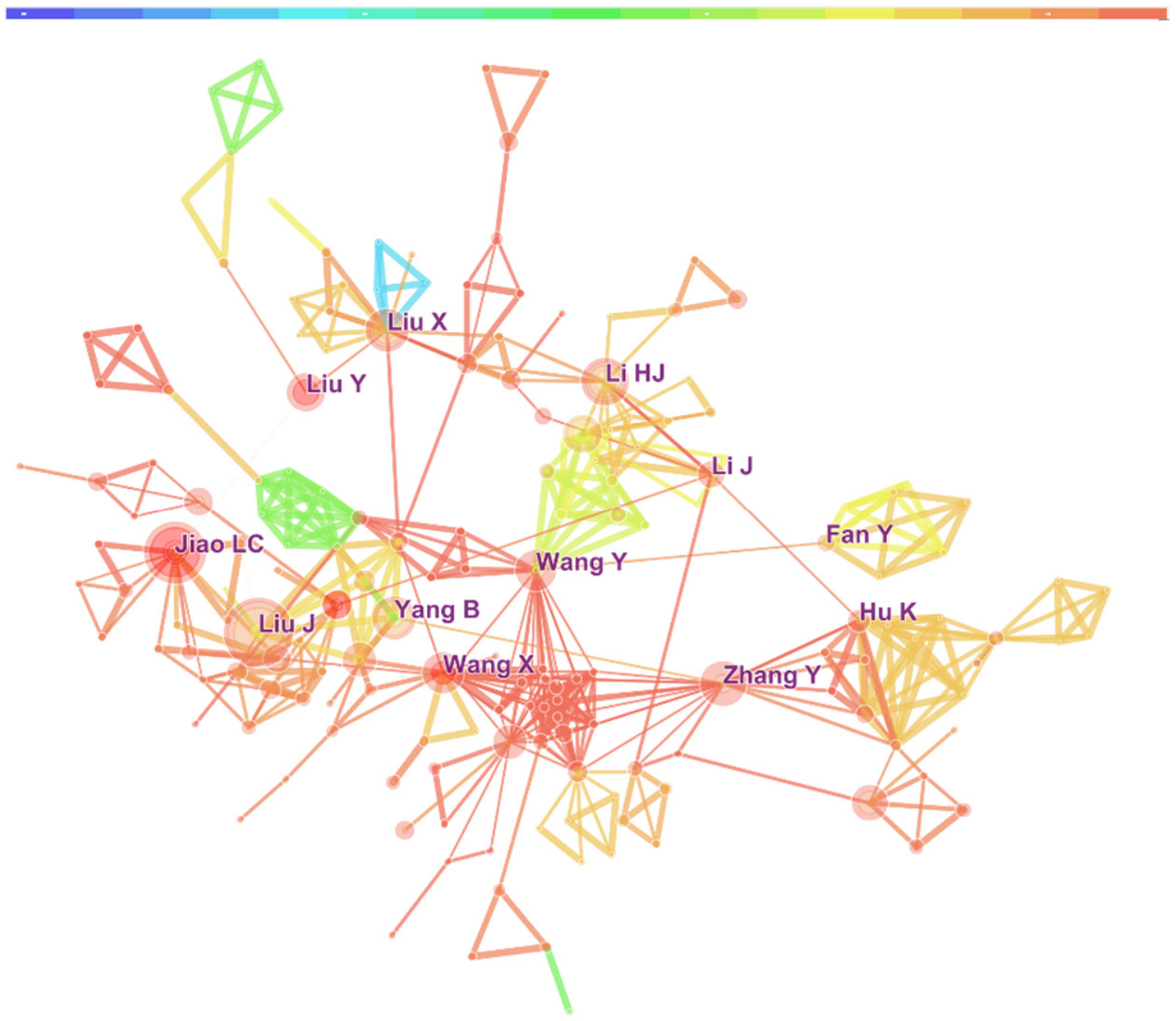

**Figure 4.** The visualisation of collaborative author network in "network community detection" literature for years 1991 to 2016. The merged network contains 1414 nodes and 2534 links. Red circles indicate citation burst. Concentric circles represent temporal patterns of a single year. Links represent co-citations and colours of the links corresponds to a given time period.

Table 4 presents top five authors of the domain in terms of frequency of joint publications. J Liu has a publication frequency of 18 in the dataset of "network community detection", which is 0.568% of total records. Following J Liu is Licheng Jiao from the "Xidian University, China" with 15 publications, which is 0.473% of the total. He has total 24738 citations and 63 h-index on Google scholar. His areas of interest include machine learning, artificial intelligence, and pattern recognition. The last three authors of the domain, Hui-Jia Li, Bo Yang, and X Liu have the same frequency of 12 publications, that is 0.378% of the total. The first author, Hui-Jia Li is the Assoc. Prof. in the "School of Management Science and Engineering, Central University of Finance and Economics, China." He has total 116 citations and 7 h-index in Google scholar. The second author, Bo Yang is the Prof. in the "School of Computer Science and Technology, Jilin University, China." He has authored 90 publications. His current areas of interests are the social computing and knowledge engineering, complex/social network, data mining, and self-adaptive and self-organized multi-agent systems.

**Table 4.** The top 5 co-authors based on the frequency of joint publications in "network community detection" bibliographic data for the years 1991 to 2016. J Liu is the most productive author with the highest frequency of publications, whereas Hui-Jia Li, Bo Yang, and X Liu have the lowest frequency of 12.

| Author | Full Name | Frequency of Publications | % of 3168 Records |
|---|---|---|---|
| Liu J | J Liu | 18 | 0.568% |
| Jiao LC | Licheng Jiao | 15 | 0.473% |
| Li HJ | Hui-Jia Li | 12 | 0.378% |
| Yang B | Bo Yang | 12 | 0.378% |
| Liu X | X Liu | 12 | 0.378% |

Table 5 demonstrates top four authors in terms of betweenness centrality. It is interesting to note that the top four authors of the domain are the most central authors with a centrality score of 0.01. First is X Liu. Second is Yu Zhang, an applied scientist in the "Microsoft." He has 355 citations and an h-index of 9 on Google scholar. His areas of interest include Network Economics, Machine Learning, Social Network Analysis, and Game Theory. Third is Y Wang. The fourth is Dongxiao He, affiliated with the "College of Computer Science and Technology, Jilin University, China."

**Table 5.** Top 4 authors based on betweenness centrality in "network community detection" literature for the years 1991 to 2016. Liu X, Zhang Y, Wang Y, and He DX are the most central authors with the highest betweenness centrality score of 0.01.

| Author | Author's Full Name | Betweenness Centrality |
|---|---|---|
| Liu X | X Liu | 0.01 |
| Zhang Y | Y Zhang | 0.01 |
| Wang Y | Y Wang | 0.01 |
| He DX | Dongxiao He | 0.01 |

Table 6 lists top five authors based on citation burst. We can see that Licheng Jiao has strongest citation burst of 4.42. Following him is Y Liu with a citation burst of 3.79. Next is X Wang with citation burst 7.14. Following him is J Liu with a citation burst of 3.38. Finally, we have Charles M Gray with a citation burst of 3.32. He is a Prof. of Cell Biology and Neuroscience. His area of interest is the Neural basis of perception.

**Table 6.** Top five authors based on the citation burst in "network community detection" bibliographic data for the years 1991 to 2016. Jiao LC has strongest citation burst, which took place in 2013. Whereas GRAY CM has lowest citation burst, which took place in 1993.

| Author | Author's Full Name | Citation Burst | Year |
|---|---|---|---|
| Jiao LC | Licheng Jiao | 4.42 | 2013 |
| Liu Y | Y Liu | 3.79 | 2015 |
| Wang X | X Wang | 3.51 | 2012 |
| Liu J | J Liu | 3.38 | 2009 |
| GRAY CM | Charles M Gray | 3.32 | 1993 |

It is pertinent to note that like other bibliographic databases, WoS database also suffers from author name ambiguity. If two or more name instances share same last name and same initials of the first name then those names can be assumed to refer to one author. In our dataset Jian Liu, Jing Liu, Liu Ji-Cheng, Jason Liu, etc., are assumed as one author J Liu; similarly, Xiaohui Liu, Xiao-Jun Liu, Liu Xu, etc., are assumed referred as X Liu, likewise, Yu Wang, Yong Wang, Yang Wang, Ya Wang, etc., are identified as Y Wang.

Initial based disambiguation of author names is a commonly used approach to preprocessing bibliometric data; however, it insufficiently solves the problem. Many scholars found it misidentifying authors when relying on initial based disambiguation (J. Kim et al. 2014).

#### 4.1.1. Identification of Largest Clusters in Author Collaborative Network

For easy understanding of the domain, we have divided the author collaborative network into clusters based on representative terms. Then clusters are labelled in descending order, such that the largest cluster is represented by #0, the second largest is represented by #1, and so on. Cluster labels characterise the nature of a cluster; generated by selecting the most representative terms extracted from titles, abstracts, or index terms of the articles citing the cluster members. They represent the context in which they are cited.

Modularity and silhouette are the key metrics which play important role in understanding the structural properties of the network.

- "The *modularity* of a network measures the extent to which a network can be decomposed into multiple components or modules. This metric provides a reference of the overall clarity of a given decomposition of the network" (Chen 2014).
- "The *silhouette* value of a cluster measures the quality of a clustering configuration. Its value ranges between −1 and 1" (Chen 2014). The higher value of silhouette score, the higher the homogeneity in the cluster.

Figure 5 depicts the visual appearance of clusters in the largest connected component of the collaborative author network, which is 12% of the entire network. The centrality refers to the position of an individual actor within the network while, centralization refers to the network as a whole. The degree centralization of the largest component is 0.124, betweenness centralization is 0.233, and closeness centralization is 0.182. The Watts-Strogatz clustering coefficient (Watts and Strogatz 1998) of largest component is 0.778 and network clustering coefficient (Transitivity) (Newman et al. 2002) is 0.595, which shows significant clustering effect in the research community. The clustering coefficient shows that it is pretty much common for two authors to collaborate if they both have collaborated with a third author.

The higher clustering coefficient and an average degree of separation indicate that authors collaboration network forms "small world" (Newman 2001b). The average number of collaborators of an author in the largest component is 5. The average distance between reachable pairs is 4.44526. The most distant vertices are Yu K and Chen GS, distance (diameter of the network) is 10. The author collaboration network has an average degree of separation approximately equal to 5. The network contains 12-cores.

It can be seen in Figure 5 that Yong Wang is the pivot node which plays the brokerage role; it exclusively connects clusters #0, #1, #2, and #31 by co-authoring with the members of these clusters. Authors in the similar positions, Xiaohui Liu, Yu Zhang, and Dongxiao He, may also play similar brokerage role in the field.

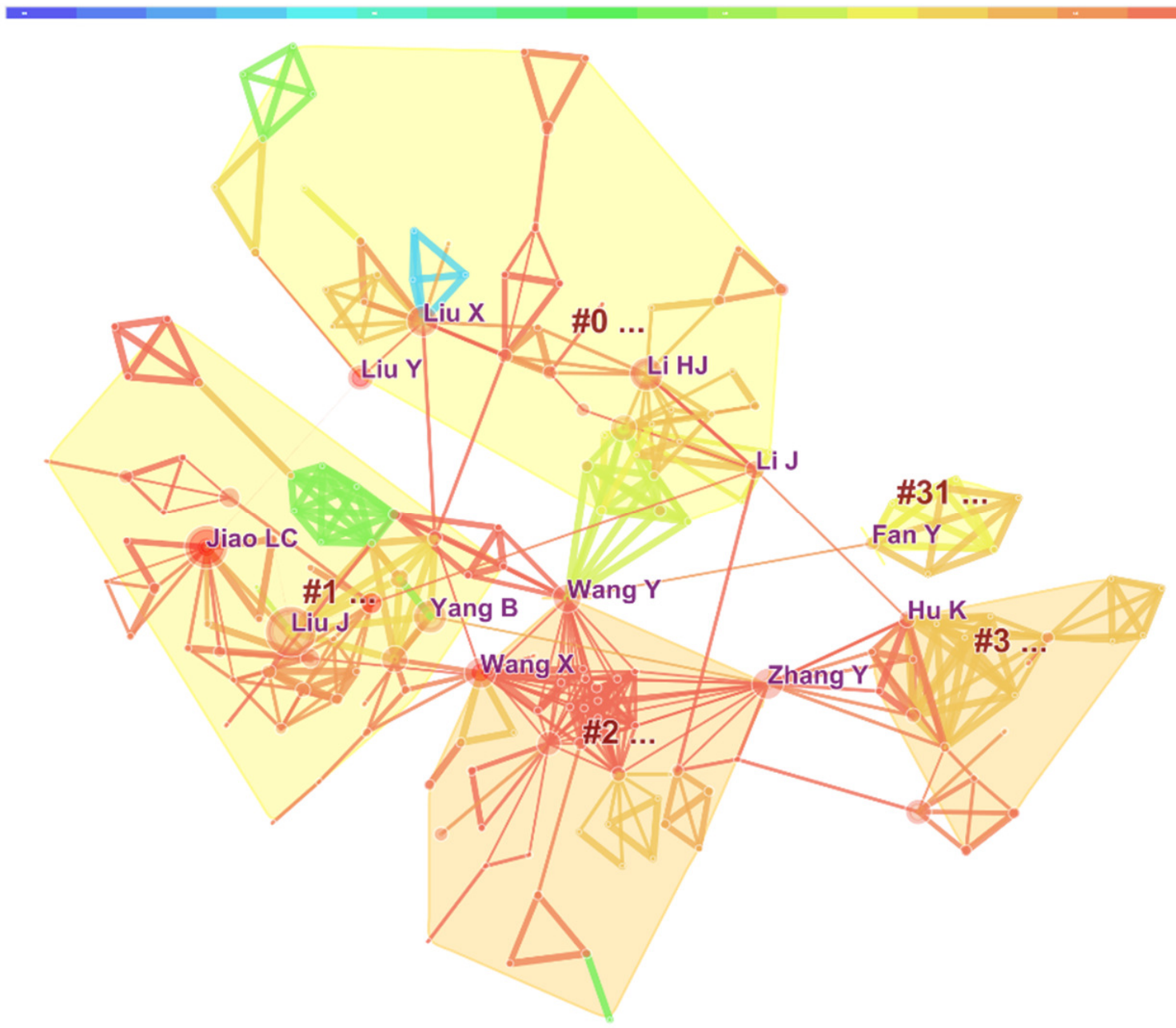

**Figure 5.** The visualisation of the giant component of the collaborative network in "network community detection" literature for the years 1991 to 2016. The largest cluster is labelled as #0, the second largest is labelled as #1, and so on. Wang Y is the pivot node. The convex hulls indicate the scope of the clusters and its colour corresponds to the average publication year of the cluster. The colour of the link indicates the time slice when the link was established first.

Table 7 demonstrates the details of the clusters of the largest connected component.

Cluster #0 (largest cluster) contains 52 nodes, which are 3.6775% of total nodes in the network. It is the major research area of the domain. The mean silhouette score is 0.964, which is the indicator of high homogeneity. The mean year of publications in this cluster is 2010.

Cluster #1 (second largest cluster) contains 51 nodes, which are 3.6067% of total nodes in the network. Cluster #1 contains high mean silhouette score of 0.954. The mean year of publications in this cluster is 2012.

Cluster #2 contains 45 nodes, which are 3.18246% of total nodes in the network. Cluster #2 contains high mean silhouette score of 0.92. The mean year of publications in this cluster is 2013.

Cluster #3 contains 21 nodes, which are 1.4851% of total nodes in the network. Cluster #3 contains high mean silhouette score of 0.982. The mean year of publications in this cluster is 2013.

Cluster #31 (smallest cluster) contains 7 nodes, which are 0.4950% of total nodes in the network. Cluster #31 contains high mean silhouette score of 0.999. The mean year of publications in this cluster is 2013.

**Table 7.** The summary table of the largest connected component of co-author network. It contains information of the cluster ID, the size of the cluster, the silhouette score indicating the homogeneity of the clusters, the average publication year of the articles in the cluster, and labels of the clusters chosen by Log-Likelihood Ratio and Mutual Information algorithms.

| Cluster ID | Size | Silhouette | Mean (Year) | Label (LLR) | Label (MI) |
|---|---|---|---|---|---|
| 0 | 52 (3.6775%) | 0.964 | 2010 | Modularity; Complex Network; Semi Supervised; | Mutation |
| 1 | 51 (3.6067%) | 0.954 | 2012 | Complex Network; Overlapping Community Detection; Incremental Method; | Bacterial |
| 2 | 45 (3.18246%) | 0.92 | 2013 | C-DBLP; Functional; Module | Non-overlapping Community |
| 3 | 21 (1.4851%) | 0.982 | 2013 | Potts Model; Time; Resolution; | Object Detection |
| 31 | 7 (0.4950%) | 0.999 | 2010 | Excitable System; Coherent Oscillation; Complex Brain Network; | Complex Network |

After giving an overview of the author collaborative network and identification of the largest cluster in this network, next we present an analysis of the author co-citation network.

## 4.2. Author Co-Citation Network Analysis

Here we demonstrate author co-citation network analysis to identify the key authors contributing to the literature of the "network community detection." The co-citation is defined as "the frequency with which two documents are cited together" (Small 1973). The merged network shown in Figure 4 contains 1382 cited authors and 2996 co-citation links. We have selected top 30 authors per one-year slice with an interval $[0, 10]$ of the citation distribution for the timespan of 1991 to 2016.

Figure 6 shows that Mark J. Newman with largest radii is the landmark node of the domain. Whereas Barabasi AL with thicker purple trim is the most central node of the domain. The red circle around Jio LC is the indicator of strongest citation burst. The detailed information can be seen in tabular form in the tables below.

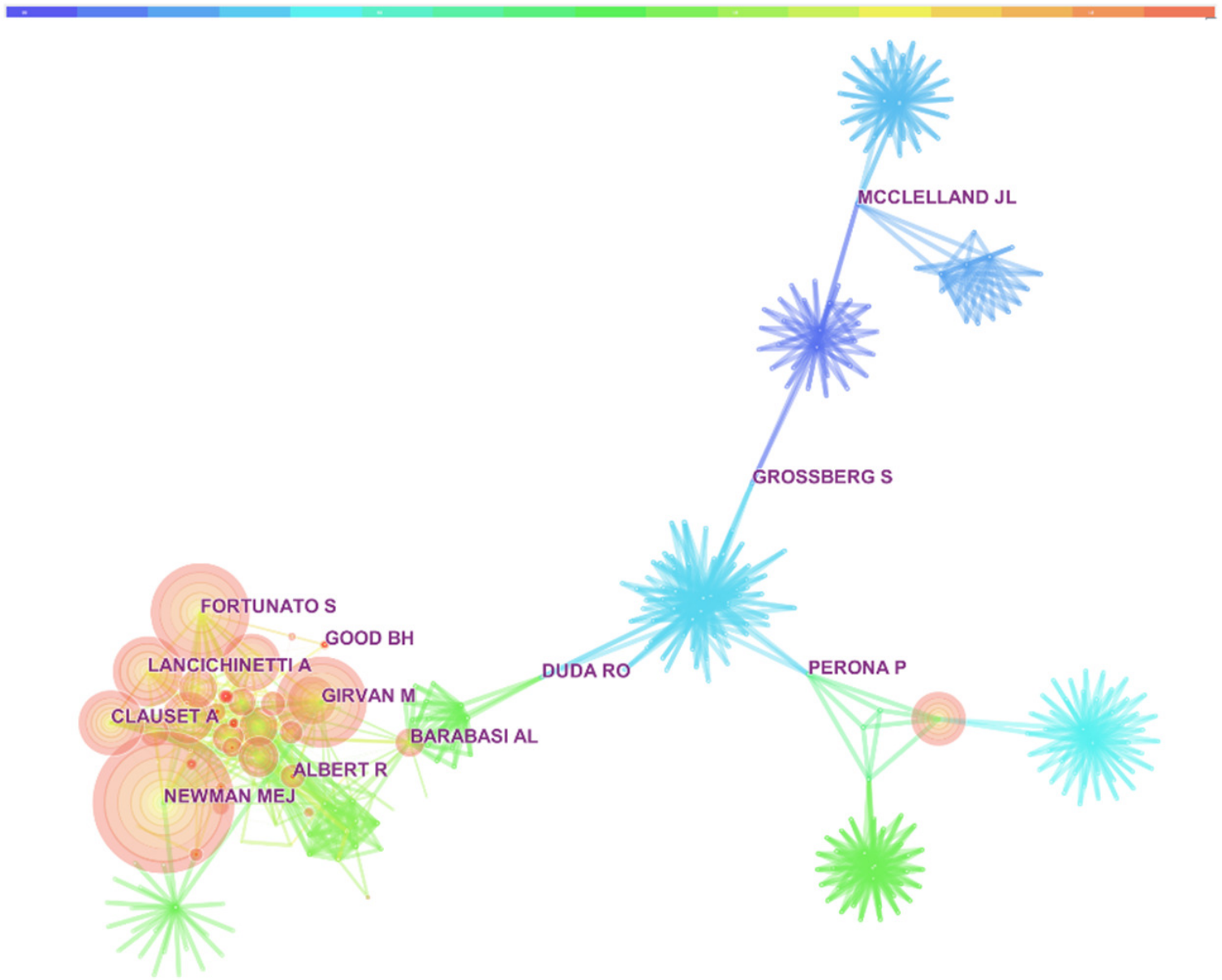

**Figure 6.** The merged network of Cited-Authors. Newman MEJ is the landmark node with largest radii. The thickness of the purple circle around Barabasi AL is the indicator of high betweenness centrality. Authors marked by red circle has strongest citation burst. The colours of the links represent the time when the link was first created.

Table 8 demonstrates the top five authors in terms of frequency of citations. Mark J. Newman has a citation frequency of 1184 in the bibliographic database of "network community detection." Newman is a Prof. of Physics in the "University of Michigan, USA." He has 246 publications and 124445 citations on Google Scholar, out of which 25 publications with 38632 citations are related to the community structure and detection (CSD). He has an h-index of 91 on Google scholar. His areas of interest include Networks and Statistical Physics. Following him is Santo Fortunato with 849 citations. He is a Prof. in the "School of Informatics and Computing, Indiana University, Bloomington, USA." He has 115 publications and 19691 citations on Google Scholar, out of which 21 publications with 13234 citations are related to the CSD. He has an h-index of 43 on Google scholar. His areas of interest include Complex Systems, Science of Science, Statistical Physics of Social Dynamics, and Networks. Next, we have Michelle Girvan with 787 citations. She is an Asst. Prof. in "University of Maryland, USA" and her areas of interest include Computational Biology and Complex Networks. She has 79 publications and 20445 citations on Google Scholar, out of which 10 publications with 18580 citations are related to CSD. She has an h-index of 17 on Google scholar. Then, we have Andrea Lancichinetti with 585 citations. He is from the "Umeå University, Sweden" and his areas of interest include Networks and Complex Systems. He has 25 publications and 5784 citations Google Scholar, out of which 13 publications with 4741 citations are in the domain of CSD. He has an h-index of 13 on Google scholar. Finally, we have Aaron Clauset with 12 citations. He is an Asst. Prof. of Computer Science in the "University of Colorado Boulder, USA." His areas of interest include Complex Networks, Network Science, Computational Social Science, Computational Biology, and Machine Learning. He has 77 publications and 14262 citations on Google Scholar, out of which 7 publications with 4851 citations are in the domain of CSD. He has an h-index of 28 on Google scholar.

It is interesting to note that the top author Newman M leads over second author Fortunato S with a major difference of 131 publications and 104754 citations on Google scholar.

**Table 8.** The first column contains top 5 authors based on the frequency of citations in "network community detection" bibliographic data for the years 1991 to 2016. It also includes citation frequency of authors, no. publications on Google Scholar, and no. of citations on Google Scholar.

| Author | Full Name | Citation Frequency | Publications on GS | Citations on GS |
|---|---|---|---|---|
| Newman MEJ | Mark J. Newman | 1184 | 246 | 124445 |
| Fortunato S | Santo Fortunato | 849 | 115 | 19691 |
| Girvan M | Michelle Girvan | 787 | 79 | 20445 |
| Lancichinetti A | Andrea Lancichinetti | 585 | 25 | 5784 |
| Clauset A | Aaron Clauset | 566 | 77 | 14262 |

Table 9 contains top five co-cited authors based on centrality. It is interesting to note that top four authors of the domain, Albert-László Barabási, Pietro Perona, Stephen Grossberg, and Richard O. Duda are the most central authors with an equal centrality score of 0.04. Whereas James L. Mcclelland with a centrality score of 0.03 is at the top fifth position

The first author Barabási is affiliated with the "Northeastern University, USA" and the "Harvard Medical School, USA." His areas of interest include Statistical Physics, Network Science, Physics, Medicine, and Biological Physics. He has 155142 citations and 117 h-index. The second author Perona P is an Allen E. Puckett Prof. in the "California Institute of Technology, USA." His areas of interest include Machine Learning, Computer Vision, Psychology, Neuroscience, and Applied Mathematics. He has 51538 citations and 85 h-index on Google scholar. The third author Grossberg S is the Wang Prof. in "Department Cognitive and Neural Systems, Boston University, USA." His areas of interest include Theoretical Cognitive Science and Psychology, Neuromorphic Technology, Computational Neuroscience, and Applied Mathematics. He has 59073 citations and 117 h-index on Google scholar. The fourth author Duda RO is an Emeritus Prof. of Electrical Engineering at the "San Jose State University, USA." His areas of interest include pattern recognition and sound localisation. The fifth author Mcclelland JL is a Prof. of Psychology in the "Stanford University, USA." His areas of interest include Cognitive Neuroscience and Cognitive Science. He has 79435 citations and 93 h-index.

**Table 9.** Top 5 cited-authors based on centrality in "network community detection" literature for the years 1991 to 2016. Barabasi AL, Perona P, Grossberg S, and Duda RO are the most central authors with the highest centrality score of 0.04. Whereas Mcclelland JL has the lowest centrality score of 0.03.

| Author | Author's Full Name | Centrality | Year |
|---|---|---|---|
| Barabasi AL | Albert-László Barabási | 0.04 | 2003 |
| Perona P | Pietro Perona | 0.04 | 1996 |
| Grossberg S | Stephen Grossberg | 0.04 | 1992 |
| Duda RO | Richard O. Duda | 0.04 | 1996 |
| Mcclelland JL | James L. Mcclelland | 0.03 | 1992 |

Table 10 lists top five co-cited authors based on burstness. We can see that Réka Albert (2003) has strongest citation burst of 23.39. She is a Distinguished Prof. in the "Pennsylvania State University, USA." Her areas of interest include Mathematics, Biology, Physics, and Networks. She has 75521 citations 46 h-index. Following her is Jussi M. Kumpula (2008) with citation burst of 21.82. He is with the "Department of Biomedical Engineering and Computational Science, Helsinki University of Technology, Finland." Next is Benjamin H Good (2010) with citation burst 21.61. He is a Miller Fellow in the "University of California, Berkeley, USA." His areas of interest include Experimental Evolution and Population Genetics. He has 698 citations and 7 h-index on Google scholar. Following him is Erzsébet Ravasz (2007) with a citation burst of 20.62. He is with the "Department of Physics, 225 Nieuwland Science Hall, University of Notre Dame, Notre Dame, Indiana, USA." Finally, we have Elizabeth A. Leicht (2010) with a citation burst of 14.15. She is a Senior Research Fellow at the "University of Oxford, UK." Her areas of interest include Networks, Resilience, Risk, Financial Systems, and Critical Infrastructure. She has 2218 citations and 9 h-index on Google scholar.

**Table 10.** Top 5 cited authors based on the citation burst in "network community detection" bibliographic data for the years 1991 to 2016. Albert R has strongest citation burst, whereas Leicht EA has lowest citation burst among top five authors.

| Author | Full Name | Burst | Year |
|---|---|---|---|
| Albert R | Réka Albert | 23.39 | 2003 |
| Kumpula JM | Jussi M. Kumpula | 21.82 | 2008 |
| Good BH | Benjamin H Good | 21.61 | 2010 |
| Ravasz E | Erzsébet Ravasz | 20.62 | 2007 |
| Leicht EA | Elizabeth A. Leicht | 14.15 | 2010 |

#### 4.2.1. Identification of Largest Clusters in Author Co-Citation Network

The author co-citation network shown in Figure 7 is decomposed into 148 co-citation clusters. These clusters are labelled by noun phrases extracted from index terms of their own citers. The 7 clusters of the largest connected component (12% of the entire network) are summarised in Table 11. It is interesting to note that the largest cluster (#0) is not part of the largest component.

There are several pivot points in the largest component which play brokerage role between different clusters, such as Barabási AL joins cluster #2 and cluster #25, Duda RO joins cluster #1 and cluster #25, Perona P joins cluster #1 and #7, Grossberg G joins cluster #1 and cluster #5, Mcclelland JL joins cluster #9 and cluster #5. It is pertinent to note here that WoS identifies a pivot node labelled as Anonymous which joins cluster #6 and cluster #7.

A cluster is a subgraph of the network. In this context, a speciality is the underlying grouping that the cluster is representing. The author co-citation analysis aims at identifying underlying specialities/clusters in a domain in terms of the groups of co-cited authors in relevant literature.

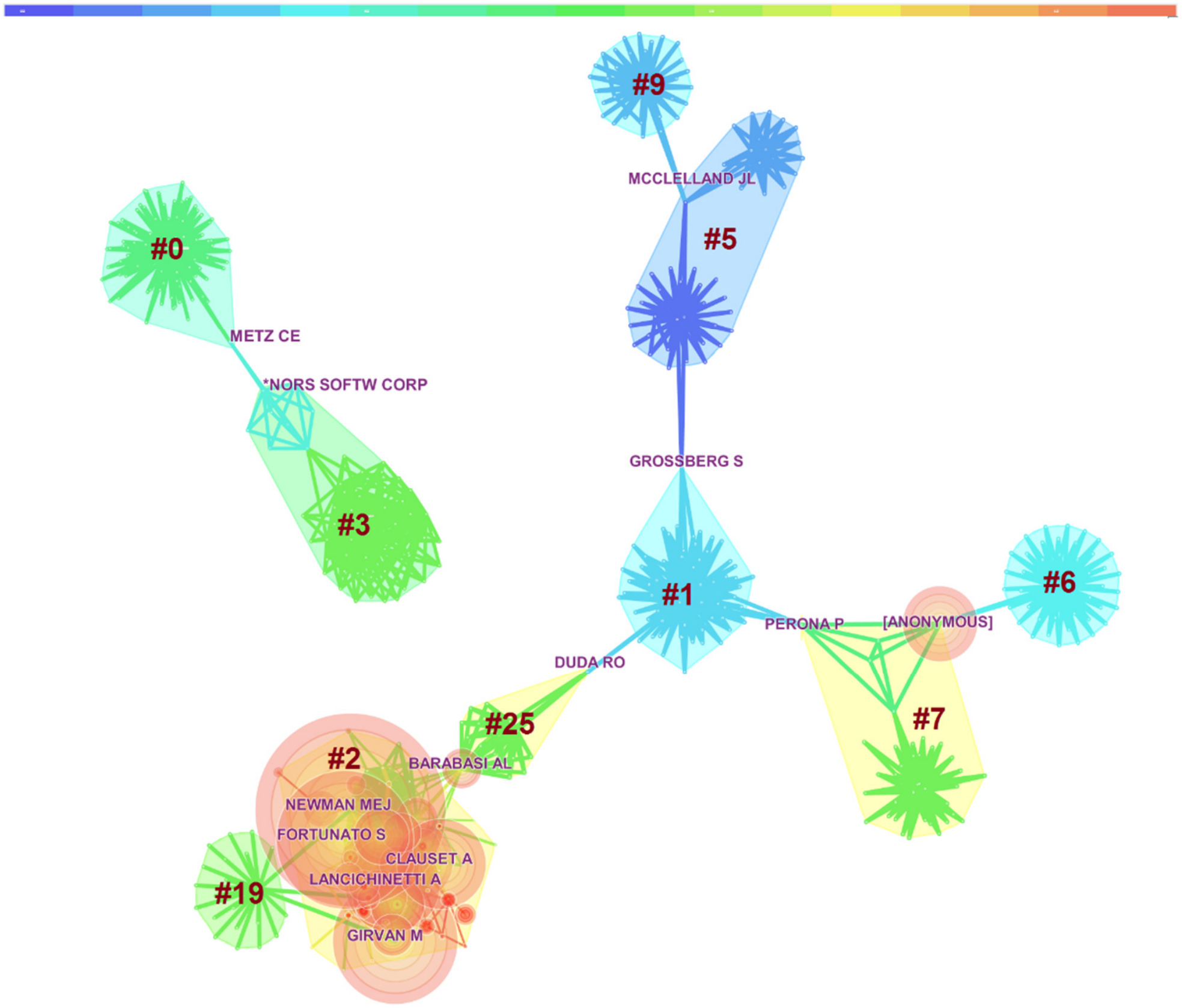

**Figure 7.** The visualisation of the giant component of the author co-citation network in the literature of "network community detection" for the years 1991 to 2016. The largest component is decomposed into 7 clusters. The clusters are labelled in descending order. The largest cluster is the cluster #0.

The largest cluster/speciality (#0) has 85 members, which is 0.061% of the entire network and a silhouette value of 0.987. It is labelled as digital radiography by LLR and neuroscience by MI. The most active citer to the cluster is Paul Sajda (2002) "multi-resolution and wavelet representations for identifying signatures of disease." It cited 99.0% of cluster members.

The second largest cluster/speciality (#1) has 60 members, which is 0.043% of the entire network and a silhouette value of 0.982. It is labelled as robot vision by LLR and zebrafish by MI. The most active citer to the cluster is Bahram Parvin (1996) "B-rep object description from multiple range views." It cited 97.0% of cluster members.

The third largest cluster/speciality (#2) has 59 members, which is 0.042% of the entire network and a silhouette value of 0.927. It is labelled as a complex network by LLR and dimensionality reduction by MI. The most active citer to the cluster is Filippo Radicchi (2010) "combinatorial approach to modularity." It cited 87.0% of cluster members.

**Table 11.** The summary table of the largest cluster and clusters of a largest connected component of author co-citation network. It contains information of the cluster ID, the size of the cluster, the silhouette score, the average citee year of the articles in the cluster, and labels of the clusters chosen by Log-Likelihood Ratio and Mutual Information algorithms.

| Cluster ID | Size | Silhouette | Mean (Citee Year) | Label (LLR) | Label (MI) |
| --- | --- | --- | --- | --- | --- |
| 0 | 85 (0.061%) | 0.987 | 2001 | Digital Radiography | Neuroscience |
| 1 | 60 (0.043%) | 0.982 | 1995 | Robot Vision | Zebrafish |
| 2 | 59 (0.042%) | 0.927 | 2007 | Complex Network | Dimensionality Reduction |
| 5 | 44 (0.0318%) | 0.981 | 1992 | Naval Research Contribution | Neuroscience |
| 6 | 43 (0.311%) | 0.999 | 1997 | England | Family |
| 7 | 38 (0.027%) | 0.976 | 2004 | Novelty Detection | Flow |
| 9 | 33 (0.023%) | 0.998 | 1995 | Chronic Schizophrenia | Depression |
| 19 | 21 (0.015%) | 0.997 | 2005 | Local Flow Betweenness Centrality | Web |
| 25 | 17 (0.012%) | 0.977 | 2004 | Networked Community | Complex Network |

After visualisation author co-citation network next, we will demonstrate analysis of journal co-citation network.

### 4.3. Journal Co-Citation Network Analysis

In this section, we focus on visualising journal co-citation network for the identification of interrelated core journals in the literature of "network community detection." To build this network, we have selected one-year per slice length for the timespan of 1991 to 2016. Node selection per slice is based on the g-index with a scaling factor $k = 5$. The network in Figure 8 contains 321 co-cited journals and 1186 co-citation links.

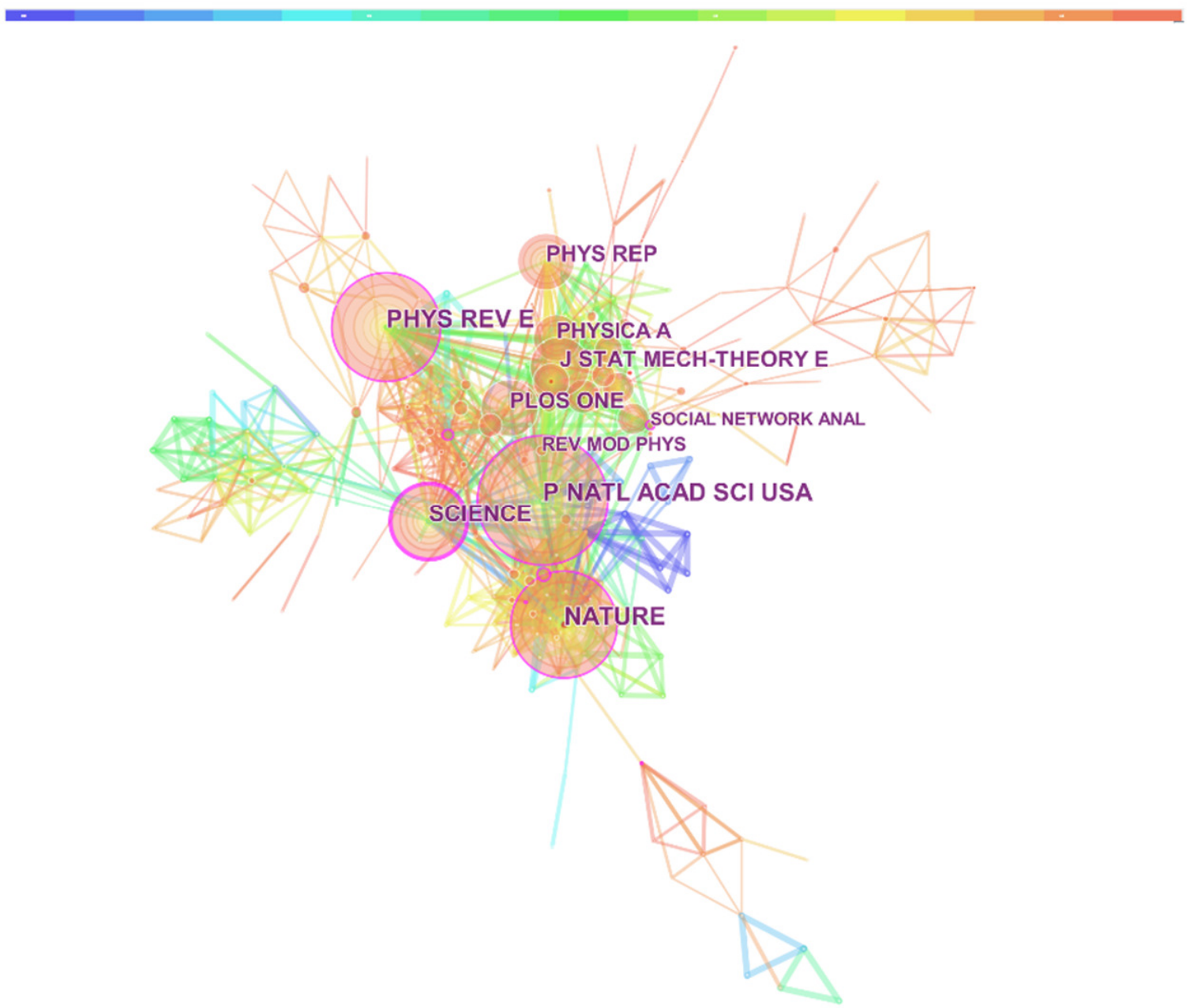

**Figure 8.** The visualisation of journal co-citation network in "network community detection" literature dataset for the years 1991-2016. The merged network contains 321 nodes and 1186 links. Concentric tree rings represent the temporal pattern of the citations of the publications in a journal. The colours of the concentric circles represent the citations in a single year, whereas colours of the links correspond to the time slice. The pink circle around the node indicates the centrality score >= 0.1.

An edge between two cited journals represents co-citation link. The thickness of a line indicates the co-citation strength. The link colour corresponds to the time slice when the link was first established between a pair of journals.

Figure 8 shows that "Proceedings of National Academy of Sciences of USA (PNAS)" with largest radii is the leading journal worldwide in the field of "network community detection." The node size is proportional to the journal's publication frequency. The concentric rings around the journal indicate the temporal citation history of the publications. The pink circle around "SCIENCE" suggests that it has centrality score >= 0.1. The red inner rings around "Review Modern Physics" signifies a rapid increase in the publications of the journal. This can be observed in tabular form as demonstrated in the tables below.

Table 12 lists the top five journals based on citation burst, suggesting a rapid increase in the number of publications. We can observe that the "Reviews of Modern Physics" has strongest citation burst of 32.64, which took place in 2004. It has 33.177 impact factor. Following it is the "JAMA: The Journal of the American Medical Association" with a citation burst of 24.85, which took place in 1997. It has 37.684 impact factor. Next is the "Science" with a citation burst of 24.74, which begin in 1992. It has 34.661 impact factor. Subsequently, we have the "The European Physical Journal B" with a citation burst of 23.89, which took place in 2004. It has 1.223 impact factor. Finally, we have the "IEEE Computer" with a citation burst of 20.37, which took place in 2004. It has 1.115 impact factor.

**Table 12.** Top five cited journals w.r.t. citation burst for "network community detection" bibliographic data for the years 1991 to 2016. Rev MOD PHYS has the strongest citation burst of 32.64, which took place in the year 2004. The Science has the highest impact factor in 2016.

| Journal | Title | 2016 Impact Factor | Burst | Year |
|---|---|---|---|---|
| Rev MOD PHYS | Reviews of Modern Physics | 33.177 | 32.64 | 2004 |
| Jama-j AM MED ASSOC | JAMA: The Journal of the American Medical Association | 37.684 | 24.85 | 1997 |
| SCIENCE | Science | 34.661 | 24.74 | 1992 |
| Eur PHYS J B | The European Physical Journal B (EPJ B) | 1.223 | 23.89 | 2004 |

| Ieee COMPUT | IEEE Computer | 1.115 | 20.37 | 2004 |
|---|---|---|---|---|

Table 13 lists details of top five most central journals based on high betweenness centrality. In terms of centrality, “Science” is the most central journal with the centrality of 0.29 and 34.661 impact factor. Following it is the “Nature” with centrality score 0.15 and 38.138 impact factor. Next, we have the “Physical Review E” with a centrality score of 0.13 and 2.252 impact factor. Following it closely are “PNAS” and “Applied and Environmental Microbiology” with centrality score 0.2 and impact factor 9.423 and 3.823 respectively.

**Table 13.** Top five cited journals w.r.t. betweenness centrality in “network community detection” bibliographic data for the years 1991 to 2016. The “Science” has the highest centrality score of 0.29, whereas “PNAS” and “Applied and Environmental Microbiology” have the lowest centrality score of 0.12.

| Journal | Title | Year | Centrality | 2016 Impact Factor |
|---|---|---|---|---|
| SCIENCE | Science | 1992 | 0.29 | 34.661 |
| NATURE | Nature | 1992 | 0.15 | 38.138 |
| Phys REV E | Physical Review E | 2004 | 0.13 | 2.252 |
| P NATL ACAD SCI USA | Proceedings of The National Academy of Sciences of USA (PNAS) | 1992 | 0.12 | 9.423 |
| Appl ENVIRON MICROB | Applied and Environmental Microbiology | 1998 | 0.12 | 3.823 |

Table 14 lists the top five most influential journals of the domain based on the number of publications. The “PNAS” with publication frequency of 1677 is identified as the key journal in the domain. It has 9.423 impact factor. Following it is the “NATURE” with publications frequency of 1427 and 38.138 impact factor. Next, we have the “Physics Review E” with a frequency of 1364 publications and 2.252 impact factor. Then, we have the “SCIENCE” with 1036 publication and 34.661 impact factor. Subsequently, we have “Journal of Statistical Mechanics: Theory and Experiment (JSTAT)” with 802 publication and 2.091 impact factor.

**Table 14.** Top five cited journals w.r.t. the frequency of publications in “network community detection” bibliographic data for the years 1975 to 2016. PNAS is the most important journal with the highest frequency of publication. Nature has the highest impact factor.

| Journal | Title | Frequency of Publication | Year | 2016 IF |
|---|---|---|---|---|
| P NATL ACAD SCI USA | Proceedings of The National Academy of Sciences of USA (PNAS) | 1677 | 1992 | 9.423 |
| NATURE | Nature | 1427 | 1992 | 38.138 |
| Phys REV E | Physical Review E | 1364 | 2004 | 2.252 |
| SCIENCE | Science | 1036 | 1992 | 34.661 |
| J STAT MECH-THEORY E | Journal of Statistical Mechanics: Theory and Experiment (JSTAT) | 802 | 2006 | 2.091 |

In the subsection below, we decompose the journal co-citation network into clusters to comprehend the structural organisation of the network.

#### 4.3.1. Identification of The Largest Cluster in Journal Co-Citation Network

This subsection identifies the largest cluster in the journal co-citation network. The synthesised network contains 530 cited journals and 2122 co-citation links. We have selected nodes per one-year slice based on the g-index with scaling factor $k = 10$ for the period of 1991 to 2016. Figure 9 displays the clusters in the giant component of the journal co-citation network. The giant component contains a total of 430 journals, including three pivot points. Giant component is 81% of the total nodes in the network.

It can be seen in Figure 9 that “Environmental Microbiology” and “Journal of Clinical Microbiology” are the pivot nodes which joins cluster #2 and cluster #4. The “ANN Internal Medicine” is also a pivot point which joins cluster #6 with cluster #4 and cluster #1.

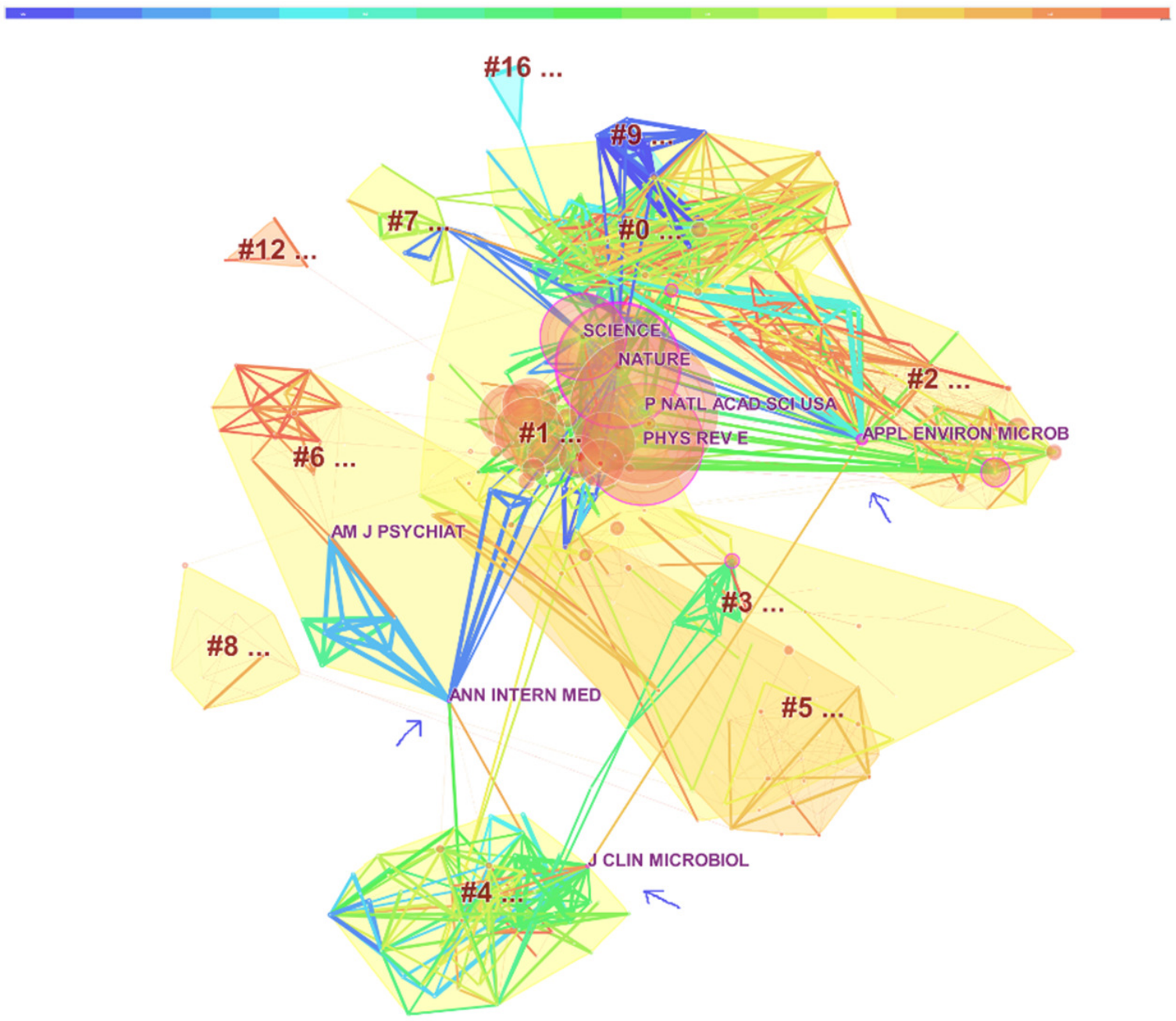

**Figure 9.** The visualisation of the largest connected component of the journal co-citation network in the literature of "network community detection" for the years 1991 to 2016. The colour of the convex hulls corresponds to the average publication year of the cluster. The colour of the node corresponds to the colour of the cluster. The red circles highlight the citation burst. Clusters are labelled in descending order based on the size.

Table 15 contains the detailed analysis of the clusters of the largest connected component in journal co-citation network.

Cluster labels characterise the nature of a cluster; they are extracted from titles, abstracts, or index terms of the articles citing the cluster members. They represent the context in which they are cited.

Cluster #0 (largest cluster) labelled as *Biodiversity; Diversity; Ecological Network*, contains 82 cited journals, which are 15.471% of total journals in the network. It is also labelled as *Adaptation* by MI. These labels represent the context in which the articles cite the cluster members. Cluster #0 is the major research area of the domain. The mean silhouette score is 0.786, which is the indicator of relatively high homogeneity. The average year of publications in this cluster is 2005.

Cluster #1 labelled as *Complex Network; Modularity; Community Structure,* contains 77 cited journals, which are 14.528% of total journals in the network. The mean silhouette score is 0.845, which is the indicator of high homogeneity. The average year of publications in this cluster is 2009. It is the second major and most active research area of the domain. Most of the highly cited, most central, most active journals of the domain belong to this cluster. The highly cited and most central members of this cluster include: "PNAS" with 1677 citations, 0.04 centrality, and 9.423 impact factor; "Nature" with 1427 citations, 0.37 centrality, and 38.138 impact factor; "Physics Review E" with 1364 citations, 0.10 centrality, and 2.252 impact factor; "Science" with 1036 citations, 0.15 centrality, and 34.661 impact factor; "JSTAT" with 802 citations, 0.02 centrality, and 2.091 impact factor. The members which make it most active area of research include: "Review Modern Physics" with 32.07 citation burst and 33.177 impact factor; "JAMA: The Journal of the American Medical Association" with 22.66 citation burst 7.48 impact factor; "European Physical Journal B" with 22.64 citation burst and 1.223 impact factor; "IEEE Computer" with 20.26 citation burst and 1.115 impact factor; "Science" with 19.79 citation burst and 34.661 impact factor.

It is interesting to note that top five highly cited articles, top five most central articles, and articles with strongest citation burst were published in the journals belonging to cluster #1.

Cluster #16 (smallest cluster) contains 3 cited journals, which is 0.754% of total journals in the network. The mean silhouette score is 0.845, which is the indicator of high homogeneity. The average year of publications in this cluster is 1997.

**Table 15.** The summary table of the largest connected component of journal co-citation network. It contains information of the cluster ID, the size of the cluster, the silhouette score indicating the homogeneity of the clusters, the average publication year of the articles in the cluster, and labels of the clusters based on representative terms.

| Cluster ID | Size | Silhouette | Mean (Year) | Label (LLR) | Label (MI) |
|---|---|---|---|---|---|
| 0 | 82 (15.471%) | 0.796 | 2005 | Biodiversity; Diversity; Ecological Network | Adaptation |
| 1 | 77 (14.528%) | 0.845 | 2005 | Complex Network; Modularity; Community Structure | Cytoskeleton |
| 2 | 67 (12.64%) | 0.899 | 2008 | Diversity; Microbial Community; Sequence | Flora |
| 3 | 53 (10%) | 0.89 | 2008 | Neural Network; Recognition; Segmentation; | Temporal Rule Extraction |
| 4 | 51 (9.622%) | 0.973 | 2003 | Surveillance; Infection; Community Acquired Pneumonia; | Clinical Prediction |
| 5 | 33 (6.226%) | 0.968 | 2012 | Community Detection; Link Mining; Community Evolution; | Core Periphery Structure |
| 6 | 26 (4.905%) | 0.984 | 2005 | Functional Connectivity; Schizophrenia; FMRI; | Electroencephalogram(EEG) |
| 7 | 12 (2.264%) | 0.979 | 2003 | Boundary Layer; Digital Elevation Model; Sea Surface Salinity; | Hyperspectral Imagery |
| 8 | 11 (2.075%) | 0.97 | 2011 | Wireless Sensor Network; Mobile Social Network; Delay Tolerant Network; | Integer Linear Program |
| 9 | 11 (2.075%) | 0.913 | 1991 | Anolis Lizard; Character Displacement; Field Experiment; | Complex Network |
| 12 | 4 (0.754%) | 0.994 | 2014 | Genetic Algorithm; Signed Social Network; Evolutionary Algorithm; | Incremental Computing |
| 16 | 3 (0.566%) | 0.996 | 1997 | Rhizosphere; Biomass; Elevated Co2; | Fatty Acid |

After a detailed analysis of the journal co-citation network, in the next section, we will visualise document co-citation network.

### 4.4. Document Co-Citation Network Analysis

In this section, document co-citation analysis is performed for the identification of key documents in the timespan of 1991 to 2016. Node selection is based on the g-index with scaling factor $k = 5$ for each slice of length one-year. Each year slice is represented by a unique colour. Link colour also corresponds to the particular time slices. The merged network contains 310 documents and 1118 links. "The network of scientific publications, linked by citations, follows a power law" (Redner 1998; B. A.-L. Barabási and Bonabeau 2003). The network shown in Figure 10 is the largest connected component of the document co-citation network. It can be observed in Figure 10 that the largest diameter of Fortunato S (2010) is the indicator of the highest frequency of citations. The thickness of purple trims around Lancichinetti A (2008) indicates that it is the most central node of the domain. The pink trims the nodes are the indicator of the centrality score $>= 0.1$.

We also observed in Figure 10 that the top articles with strongest citation burst, such as Girvan M (2002) published in "PNASP," Newman MEJ (2004) published in "Phys Rev E," Newman MEJ (2003) published in SIAM Rev," Albert R (2002) published in "Rev Mod Phys," and Clauset A (2004) published in "Phys Rev E," belong to cluster #1 of the journal co-citation network.

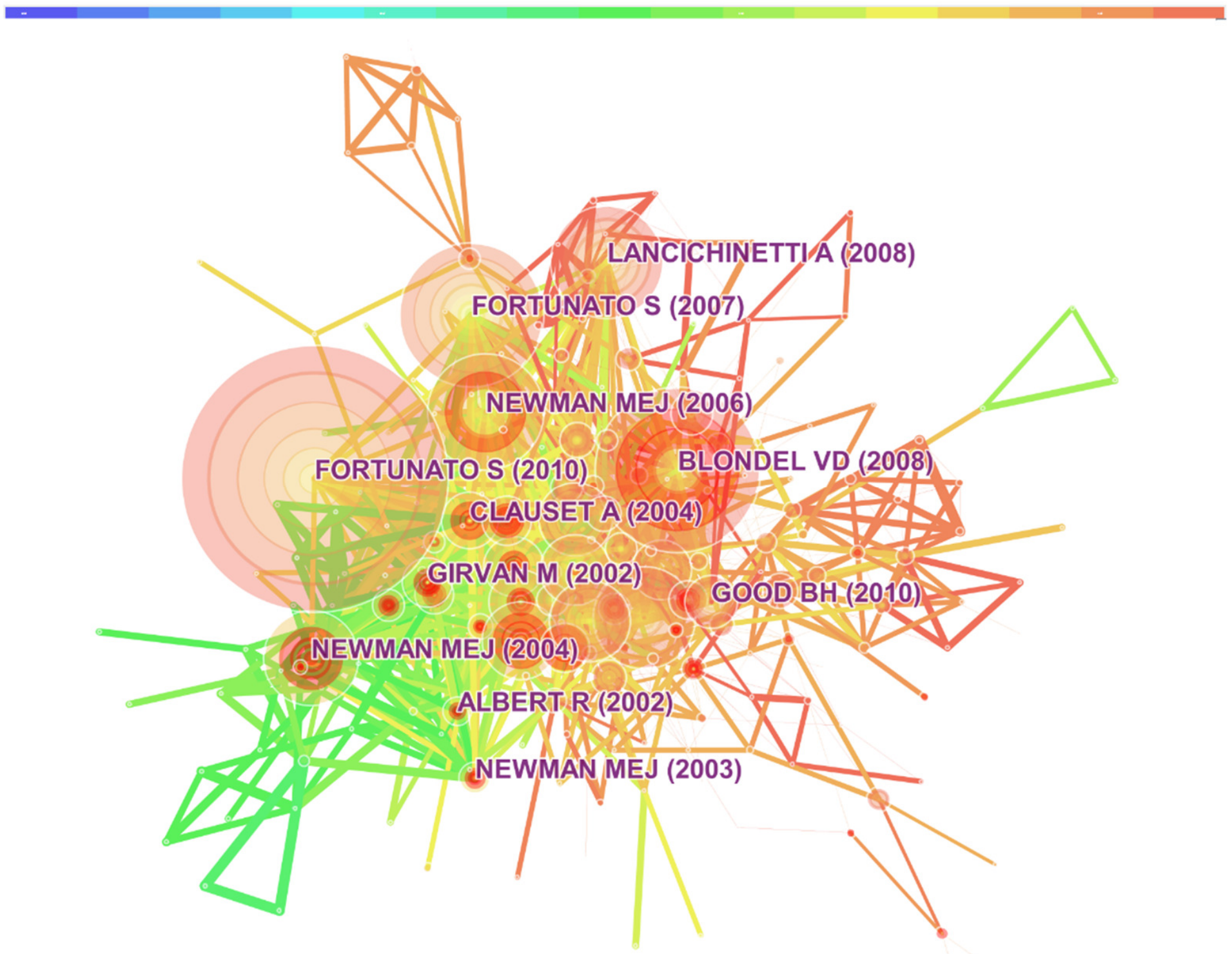

**Figure 10.** The visualisation of the document co-citation network of "network community detection" literature for the years 1991-2016. Red circles around the nodes indicate citation burst of the node. The colours of the links correspond to the particular time slice. The blue colour indicates starting years, green colour represents middle years, whereas yellowish colour represents most recent years of the given time span. The concentric tree rings represent the temporal pattern of publications.

Table 16 demonstrates the top five cited references based on citation frequency. The article by Santo Fortunato (2010), entitled "community detection in graphs" has highest citations of 652, which is 20.580% of entire records in the dataset. It was published in the "Physics Reports" and has a half-life of 4 years. It is a comprehensive survey paper of community detection methods, with 4892 citations on Google scholar.  Following it is the article by Vincent D. Blondel (2008), entitled "Fast unfolding of communities in large networks," with 422 citations which are 13.320%. It was published in "JSTAT" and has a half-life of 6 years. In this article, Blondel et al. have proposed a well-known and most efficient community detection technique called "Louvain." This article has 4721 citations on Google scholar. Next is the article by Santo Fortunato (2007), entitled "Resolution limit in community detection" with 362 citations, which is 11.426%. It was published in "PNAS" and has a half-life of 6 years. It has 1624 citations on Google scholar. In this article, Fortunato et al. have identified resolution limit problem in modularity— a well-known quality function. Subsequently, we have an article by Mark E. J. Newman (2006), entitled "Modularity and community structure in networks" with 304 citations which are 9.595%. In this article, Newman has introduced modularity matrix to improve spectral optimisation of modularity. It was published in "PNAS" and has a half-life of 6 years. It has 4790 citations in Google Scholar. Finally, we have an article by Andrea Lancichinetti (2008), entitled "Benchmark graphs for testing community detection algorithms" with 297 citations, which are 9.35%. It was published in "Physics Review E" and has a half-life of 6 years. It has 1172 citations in Google Scholar. In this article, Lancichinetti et al. have introduced LFR benchmark.

**Table 16.** The top five cited references based on frequency in "network community detection" bibliographic data for the years 1991 to 2016. The article of Fortunato S has the highest frequency of citations and the article of Lancichinetti A has the lowest frequency of citations.

| Author | Full Name | Freq. | Year | Source | Title | Vol | Page | Half-life |
|---|---|---|---|---|---|---|---|---|
| Fortunato S | Santo Fortunato | 652 | 2010 | PHYS REP | Physics Reports | V486 | P75 | 4 |
| Blondel VD | Vincent D. Blondel | 422 | 2008 | J STAT MECH-THEORY E | Journal of Statistical Mechanics: Theory and Experiment (JSTAT) | V10 | P10008 | 6 |
| Fortunato S | Santo Fortunato | 362 | 2007 | P NATL ACAD SCI USA | Proceedings of the National Academy of Sciences of USA (PNAS) | V104 | P36 | 6 |
| Newman MEJ | Mark E. J. Newman | 304 | 2006 | P NATL ACAD SCI USA | Proceedings of the National Academy of Sciences of USA (PNAS) | V103 | P8577 | 6 |
| Lancichinetti A | Andrea Lancichinetti | 297 | 2008 | PHYS REV E | Physics Review E | V78 | P046110 | 6 |

Table 17 demonstrates co-cited documents in terms of centrality. The article of Andrea Lancichinetti (2008), published in "Physics Review E" has the highest centrality score of 0.12. Following it is the article of Benjamin H Good (2010), also published in "Physics Review E" has a centrality score of 0.1. Subsequent three articles by Santo Fotunato (2010), Vincent D. Blondel (2008), and Santo Fotunato (2007) have an equal centrality score of 0.06. The article of Fortunato S (2010) was published in "Physics Reports." The article of Blondel VD (2008) was published in "JSTAT" and the article of Fortunato S (2007) was published in "PNAS."

**Table 17.** Top five cited references w.r.t. betweenness centrality (BC) in "network community detection" bibliographic data for the years 1991 to 2016. The article of Lancichinetti A (2008) published in the "Physics Review E" has the highest centrality score of 0.12. The half-life of this article is 6 years.

| Author | Author's full name | BC. | Year | Source | Title | Vol. | Page | Half-life |
|---|---|---|---|---|---|---|---|---|
| Lancichinetti A | Andrea Lancichinetti | 0.12 | 2008 | PHYS REV E | Physics Review E | V78 | P046110 | 6 |
| Good BH | Benjamin H Good | 0.1 | 2010 | PHYS REV E | Physics Review E | V81 | P046106 | 3 |
| Fortunato S | Santo Fortunato | 0.06 | 2010 | PHYS REP | Physics Reports | V486 | P75 | 4 |
| Blondel VD | Vincent D. Blondel | 0.06 | 2008 | J STAT MECH-THEORY E | Journal of Statistical Mechanics: Theory and Experiment (JSTAT) | V10 | P10008 | 6 |
| Fortunato S | Santo Fortunato | 0.06 | 2007 | P NATL ACAD SCI USA | Proceedings of the National Academy of Sciences of USA (PNAS) | V104 | P36 | 6 |

By observing Table 16 and Table 17, we identified that the top five highly cited articles and top five most central articles are published in the journals which all belong to cluster #1 of the journal co-citation network.

In the subsection underneath, we will partition the document co-citation network into clusters to demonstrate the structural organisation of the network.

#### 4.4.1. Identification of Largest Network in Document Co-Citation Network

The panoramic network contains 988 cited documents and 3404 co-citation links. We have selected time slice of length one-year for the timespan of $1991 - 2016$. Selection criteria of nodes per slice is the g-index with scaling factor $k = 20$. Figure 11 depicts the clusters in the giant component of the document co-citation network. The giant component contains 430 documents, which is 43% of the total nodes in the network.

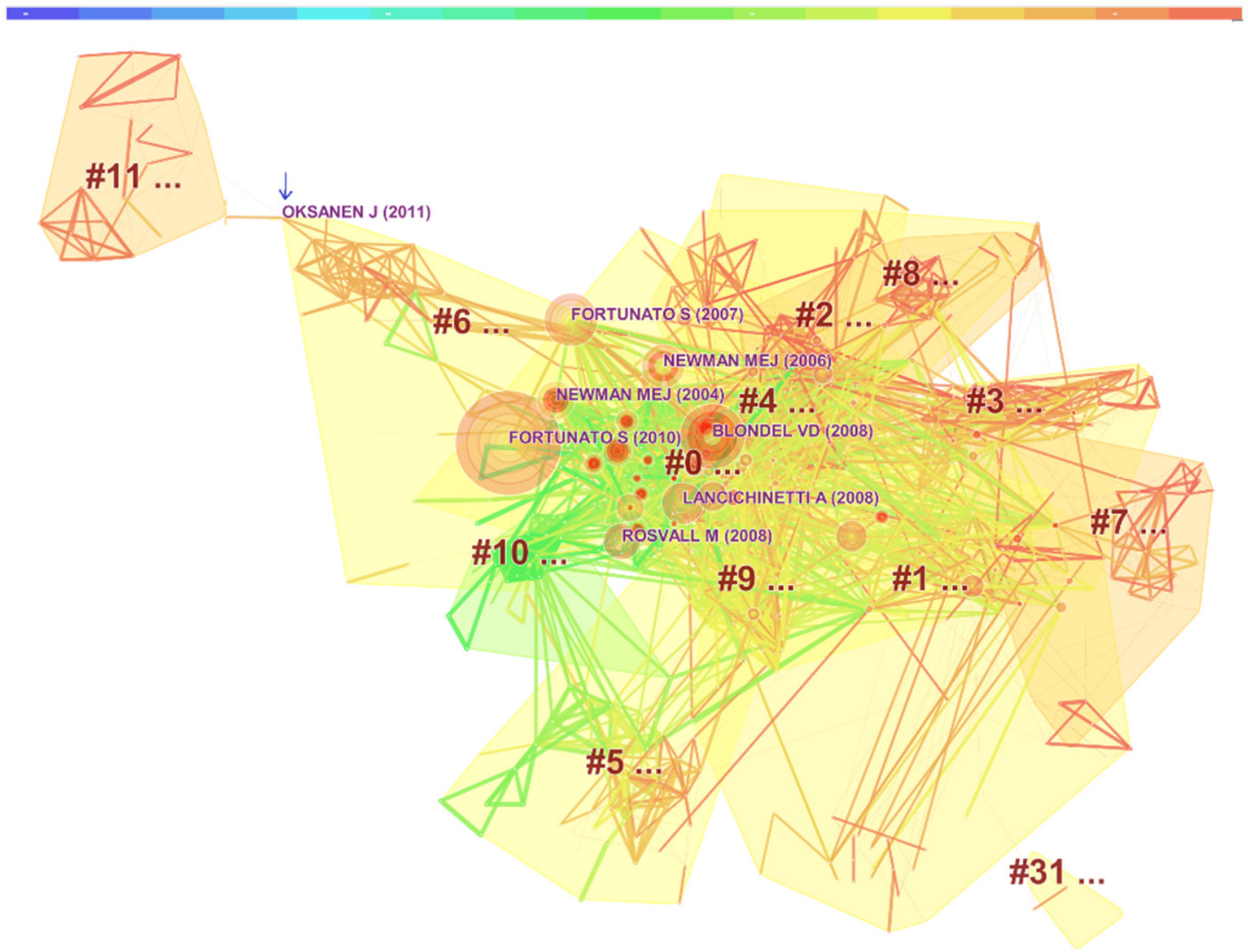

**Figure 11.** The visualisation of the giant component of the document co-citation network in "network community detection" literature for the years 1991 to 2016. Red circles highlight the strongest burstness of the nodes. Cluster #0 is the largest and most active area of the domain. Cluster #31 is the smallest cluster of the giant component.

Table 18 presents the detailed analysis of the clusters of the largest connected component of the document co-citation network. Cluster labels characterise the overall theme of a cluster; they are extracted from noun phrases of the articles citing the cluster members.

Cluster #0 is the largest cluster of the giant component, whereas cluster #31 is the smallest cluster of the connected component. The silhouette score is the indicator of the homogeneity of the clusters. The labels of the clusters are chosen by Log-Likelihood Ratio and Mutual Information algorithms. Oksanen J (2011) is the key turning point which exclusively joins cluster #6 with cluster #11.

Cluster #0 is labelled as "Modularity; Complex Network; Resolution" by LLR, and "Amphibian" by MI. It contains 99 cluster members, which are 10.020% of total documents in the network. It is the major research area of the domain. The mean silhouette score is 0.584, which is the indicator of relatively low homogeneity. The mean year of publications in this cluster is 2004. The most active citer to the cluster is Li, HJ (2012) "identifying overlapping communities in social networks using multi-scale local information expansion." It cited 8% of the cluster members. "Modularity" represents the speciality of cluster #0

As most of the documents with the highest citation burst are the members of this cluster, which indicates cluster as a whole is an active field of research. Some of the most productive, influential, and active documents belonging to this cluster are Fortunato S (2010) (Fortunato 2010), Bolondel VD (2008) (Blondel et al. 2008), Fortunato S (2007) (Fortunato and Barthelemy 2007), Lancichinetti A (2008) (Lancichinetti et al. 2008), Newman MEJ (2006) (Newman 2006), Sales-Pardo M (2007) (Sales-Pardo et al. 2007), Radicchi F (2004) (Radicchi et al. 2004), Newman MEJ (2004) (Newman 2004c), Girvan M (2002) (Girvan and Newman 2002), Clauset A (2004) (Clauset et al. 2004).

The second largest cluster (#1) is labelled as "Connectome; Functional Network; Schizophrenia;" by LLR, and "Childhood Leukaemia" by MI. It contains 77 cited documents, which are 7.793% of total documents in the network. The mean silhouette score is 0.704, which is the indicator of relatively high homogeneity. The mean publication year in this cluster is 2009. The most active citer to the cluster is 0.12 Sun, HL (2014) "IncOrder: incremental density-based community detection in dynamic networks." "Overlapping community" represents the speciality of cluster #1. It has coverage 0.12, i.e. it has cited 12% of members of clusters.

Two clusters 4 and 7 have the same labels identified by MI. As clusters are labelled by terms from citing articles, therefore, these clusters are likely cited by some papers but pulled apart enough by other papers.

**Table 18.** The summary table of the giant component of the document co-citation network. It contains information of the cluster ID, the size of the cluster, the silhouette score indicating the homogeneity of the clusters, the average publication year of the articles in the cluster, and labels of the clusters.

| Cluster ID | Size | Silhouette | Mean (Year) | Label (LLR) | Label (MI) |
|---|---|---|---|---|---|
| 0 | 99 (10.020%) | 0.584 | 2004 | Modularity; Complex Network; Resolution | Amphibian |
| 1 | 77 (7.793%) | 0.704 | 2009 | Overlapping Community; Complex Network; Overlapping Community Detection; | Childhood Leukemia |
| 2 | 64 (6.477%) | 0.762 | 2010 | Connectome; Functional Network; Schizophrenia; | Science and Technology |
| 3 | 55 (5.566%) | 0.781 | 2010 | Stochastic Blockmodel; Phase Transition; Stochastic Block Model; | Graph Limit |
| 4 | 52 (5.263%) | 0.684 | 2009 | Complex Network; Potts Model; Resolution; | Coauthor Network |
| 5 | 48 (4.858%) | 0.817 | 2005 | Saccharomyces Cerevisiae; Complex; Yeast; | Critical Phenomena of Socio-Economic System |
| 6 | 42 (4.251%) | 0.872 | 2008 | Specialization; Animal Mutualistic Network; Ecological Network; | Biodiversity Hotspot |
| 7 | 37 (3.744%) | 0.85 | 2008 | Dynamic Network; Uncertain Data Stream; Dynamic Community Detection; | Coauthor Network |
| 8 | 34 (3.44%) | 0.761 | 2010 | Genetic Algorithm; Evolutionary Algorithm; Structural Balance; | Improved EM |
| 9 | 29 (2.935%) | 0.899 | 2008 | Community Structure Detection; Resolution; Variable Neighborhood Search; | Arabidopsis |
| 10 | 27 (2.732%) | 0.92 | 2001 | Small World; Betweenness; Organization; | LGT |
| 11 | 23 (2.327%) | 0.981 | 2010 | Complex Network; Diversity; Microbial Community; | Biological Control Agent |
| 31 | 5 (0.506%) | 0.987 | 2008 | Social Circle; Community Profiling; Heterogeneous Social Network; | Complex Network |

After an overview of the analysis of the cited references, in this section, we will perform analysis of the collaborative country network.

## 4.5. Collaborative Country Network Analysis

This section presents the visualisation of the spread of research in "network community detection" from various territories. For this analysis, we have chosen top 30 countries per slice of the citation distribution. We have selected slice length = 1 year for the timespan of 1991 to 2016. The merged network contains 84 countries and 495 collaborative links.

It can be noticed in Figure 12 that the United States is the landmark node with largest radii. It provides the evidence that key articles of the domain originate from the United States. The pink circles around the United States, England, France, Spain, and Norway indicate that they have centrality score $>= 0.1$. The thickness of pink trims around the United States depicts that it is also the most central country of the domain. Additional details of the visualisation are given underneath in the tables.

The number of components in the country-country network with a minimum size equal to one is 8. The largest component contains 77 vertices, that is 91.667% of the entire network. The degree centralization of the largest component is 0.423, betweenness centralization is 0.192, and closeness centralization is 0.466. The average distance between reachable pairs is 2.17225. The most distant vertices are Nepal and Kenya, network diameter is 5. This component contains 13-cores. The mean clustering coefficient of the largest component is 0.690 and Transitivity is 0.476, which depicts relatively high clustering effect. Two countries are more likely to collaborate if they both have collaborated with a third country. The average number of collaborators of a country in the largest component is 11.740 and in entire network is 10.762.

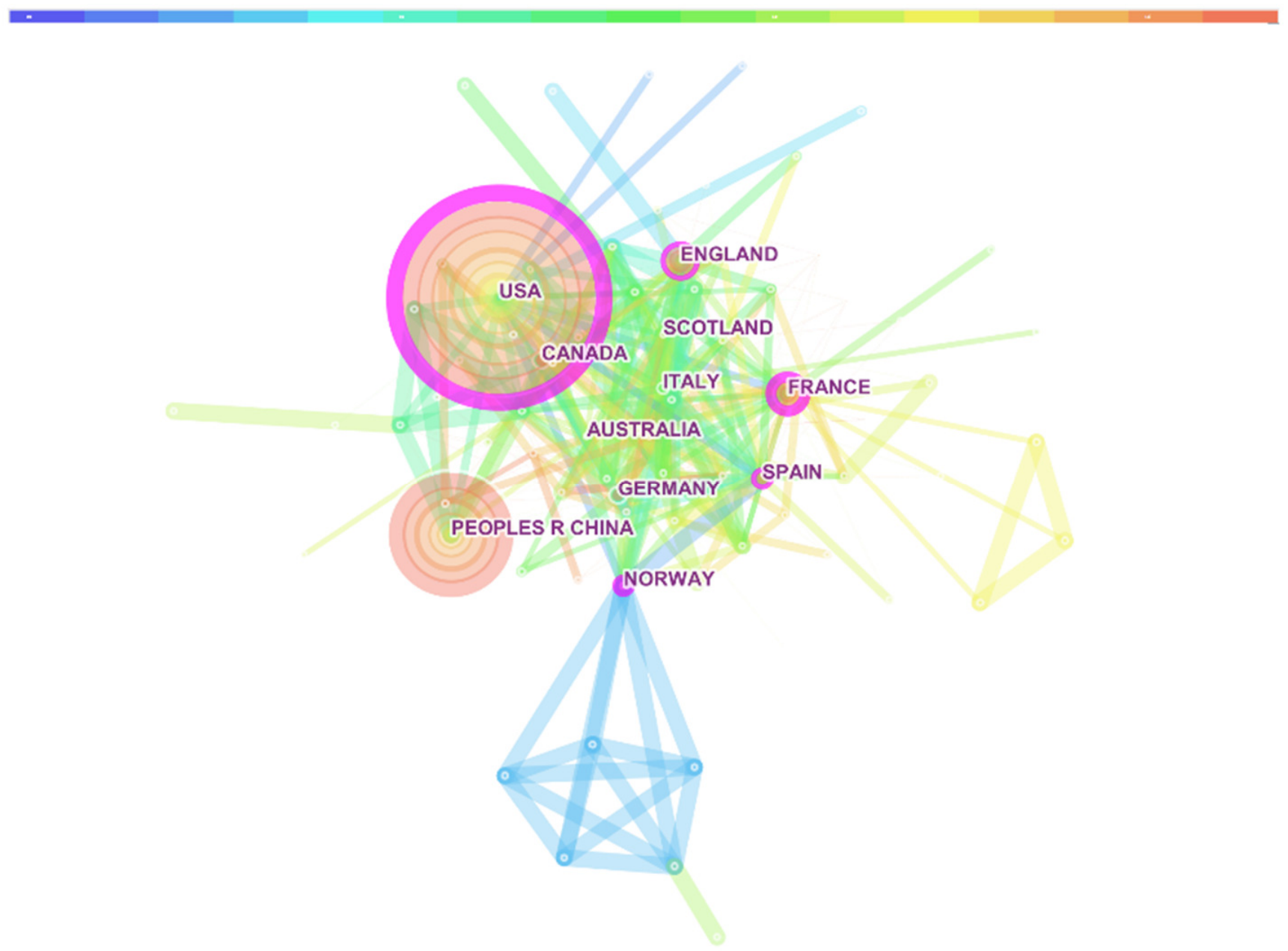

**Figure 12.** The visualisation of the network of countries in "network community detection" bibliographic data for the years 1991–2016. The merged network contains 84 countries and 495 co-authorship links. The concentric tree rings represent the temporal pattern of the publications in corresponding years. The pink circle around the nodes represent the betweenness centrality score >= 0.1. Whereas red highlighted nodes represent the strength of citation burst. The colours of the links correspond to the particular time period.

Table 19 lists top five countries in terms of betweenness centrality. The "United States" is the most influential country of the domain with a centrality score of 0.4, degree centrality of 45 and closeness centrality of 0.70. Next is the "Norway" with a centrality score of 0.24, degree centrality of 19, and closeness centrality of 0.55. It is closely followed by "France" with a centrality score of 0.24. Then we have "England" with a centrality score of 0.17. Finally, we have "Spain" with a centrality score of 0.14.

**Table 19.** The top five countries based on betweenness centrality in "network community detection" literature for the years 1991-2016. The USA is the most central country of the domain.

| Country | Betweenness Centrality | Degree Centrality | Closeness Centrality |
|---|---|---|---|
| USA | 0.4 | 45 | 0.70 |
| NORWAY | 0.25 | 19 | 0.55 |
| FRANCE | 0.24 | 35 | 0.64 |
| ENGLAND | 0.17 | 40 | 0.66 |
| SPAIN | 0.14 | 36 | 0.64 |

Table 20 contains top five countries based on the frequency of publications. The order of countries in this table is somewhat different from Table 14. The "United States" is again on the top with 1076 publications, which is 33.964% of the total records. Next, we have a fresh entry, "the Peoples' Republic of China" with 664 publications, which is 20.959%. Following it is the "England" with 267 publications, which is 8.428%. Then we have "France" with 226 publications, which is 7.133%. Subsequently, we have "Spain" with 196 publications, which is 6.1868%.

**Table 20.** The top five countries based on the frequency of publications in "network community detection" bibliographic data for the years 1991 to 2016. The USA is the most significant country of the domain with a frequency of 1076 publications. Whereas Spain has the lowest frequency of 196 publications.

| Country | Publications Frequency | % of Records | Year |
|---|---|---|---|
| USA | 1076 | 33.964% | 1992 |
| PEOPLES' REPUBLIC OF CHINA | 664 | 20.959% | 2001 |
| ENGLAND | 267 | 8.428% | 1996 |
| FRANCE | 226 | 7.133% | 1994 |
| SPAIN | 196 | 6.1868% | 1994 |

As shown in Figure 13, "Scotland" has the strongest and longest citation burst of 4.7932, which lasts for 11 years in the timespan of 1999 to 2009. It indicates that publications originating from Scotland have been the focus of attention of its scientific community. Whereas "England" has lowest citation burst and Italy has shortest citation burst of 2 years' duration.

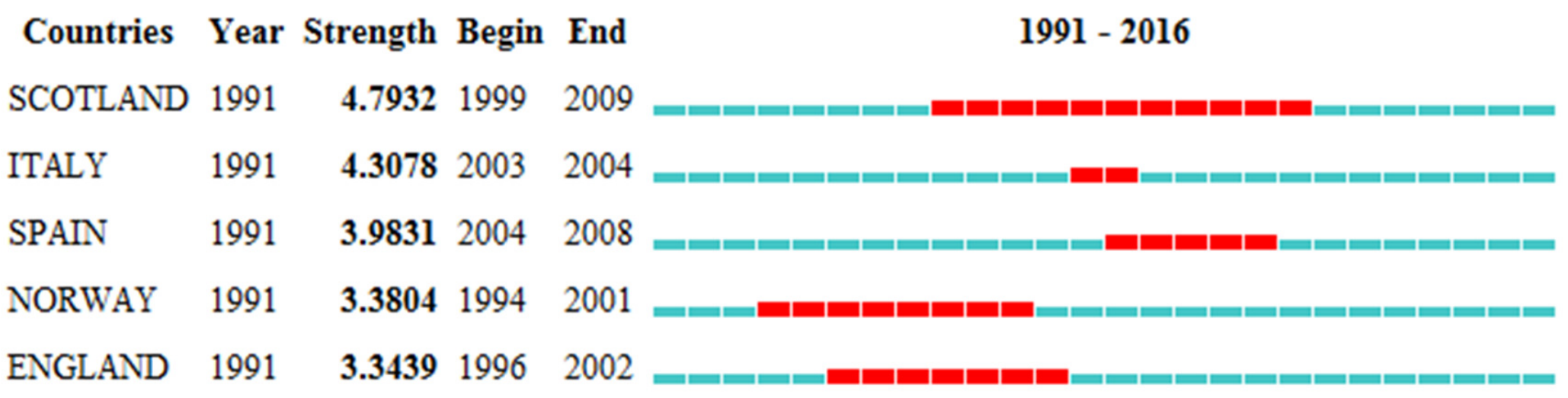

**Figure 13.** Citation burst history of countries in the domain of "network community detection" in the timespan of 1991 to 2016. The Scotland has longest and strongest citation burst, whereas Italy has shortest citation burst and England has lowest citation burst.

After analysing collaborative country network, next, we move towards visualisation of the collaborative institutions.

### 4.6. Collaborative Institution Network Analysis

This section presents the visual analysis of the collaborative institutions in the bibliographic literature of "network community detection" in the timespan of 1991 to 2016. Total records in the dataset are 3168. We have selected top 30 institutions from each time slice of length = 3. The merged network comprises of 343 institutions and 783 co-authorship links.

As depicted in Figure 14, MIT is the most central node among all other institutions. Whereas the "Chinese Academy of Sciences" is the most productive institution with the largest diameter. The purple trims around the nodes signify high betweenness centrality of the journals. Additional details are demonstrated in the tables underneath.

The number of components in the institution-institution network with a minimum size equal to two is 42. The largest component contains 147 vertices, that is 42.857% of the entire network. The average clustering coefficient of largest component is 0.744 and Transitivity is 0.558, which is the indicator of relatively high clustering effect. Two institutions are more likely to collaborate if they both have collaborated with a third institution. The degree centralization of largest component is 0.221, betweenness centralization is 0.228, and closeness centralization is 0.326. The average distance between reachable pairs is 2.278. The most distant vertices are the "University of Washington" and the "University of Wollongong," network diameter is 7. The largest component contains 13-cores. The average number of collaborators of an institution in the largest component are 8.054 and in entire network are 4.524.

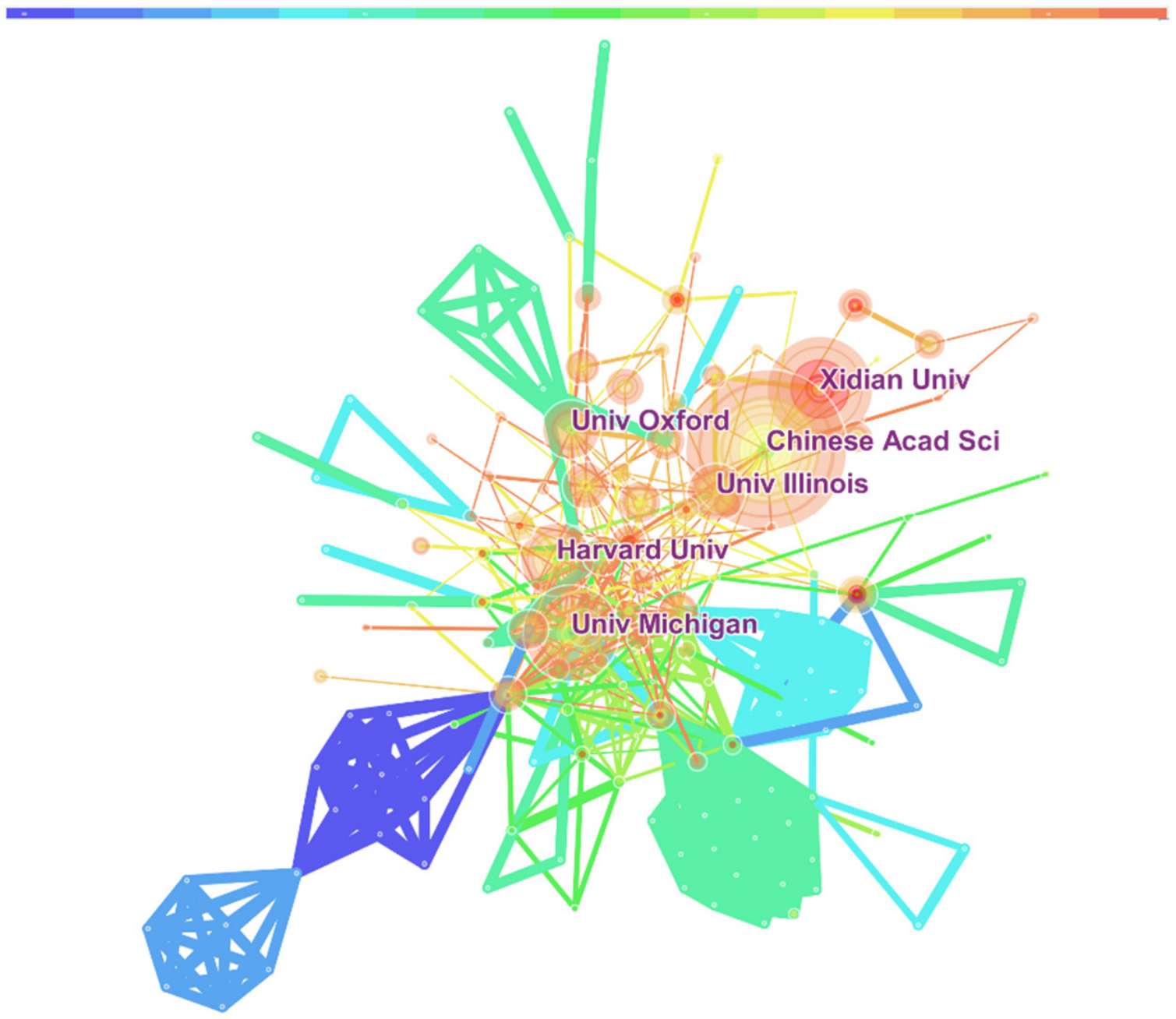

**Figure 14.** The visualisation of the merged network of institutions in the "network community detection" bibliographic data for the years 1991–2016. The merged network contains 343 organisations and 783 co-authorship links. The purple circle around the node represents the betweenness centrality score, whereas red highlighted nodes represent the strength of citation burst. The colours of the links correspond to the particular time period.

Table 21 lists top five institutions in terms of centrality. "The Massachusetts Institute of Technology (MIT), USA" is the most central institution of the domain with a betweenness centrality score of 0.09, degree centrality of 36, and closeness centrality of 0.4883. Following it is "National Aeronautics and Space Administration, USA" with a centrality score of 0.06, degree centrality of 40, and closeness centrality of 0.4695. Next, we have the "University of Oxford, UK" with a centrality score of 0.05, degree centrality of 18, and closeness centrality of 0.4160. Then we have the "University of Maryland, USA" with a centrality score of 0.05, degree centrality of 34, and closeness centrality of 0.5000. Finally, we have the "Chinese Academy of Sciences, China" with a centrality score of 0.04, degree centrality of 17, and closeness centrality of 0.3802.

**Table 21.** Top five institutions based on betweenness centrality in "network community detection" literature for the years 1975-2016. The MIT is the most central institution of the domain.

| Institution | Abbreviation | Betweenness Centrality | Degree Centrality | Closeness Centrality |
|---|---|---|---|---|
| The Massachusetts Institute of Technology | MIT | 0.09 | 36 | 0.4883 |
| National Aeronautics and Space Administration | NASA | 0.06 | 40 | 0.4695 |
| University of Oxford | Univ Oxford | 0.05 | 18 | 0.4160 |
| University of Maryland | Univ Maryland | 0.05 | 34 | 0.5000 |
| Chinese Academy of Sciences | Chinese Acad Sci | 0.04 | 17 | 0.3802 |

Table 22 presents top five institutions based on the frequency of publications in the domain of "network community detection." "Chinese Academy of Sciences, China" leads other countries with a frequency of 92 publications, that is 2.029% of entire records in the dataset. Next is "Xiadian University, China" with 61 publications, that is 1.925% of the total. Following it is the "University of Michigan, USA" with 58 publications, that is 1.830% of the total. Next, we have the "Harvard University, USA" with 46 publications, that is 1.452% of the total. Finally, we have the "University of Illinois, USA" with 34 publications, that is 1.073% of the total.

**Table 22.** Top five most productive institutions based on the frequency of publications in "network community detection" bibliographic data for the years 1991 to 2016. The "Chinese Academy of Science" is the most important institution of the domain.

| Institution | Abbreviation | Publication Frequency | % of Records | Year |
|---|---|---|---|---|
| Chinese Academy of Sciences | Chinese Acad Sci | 92 | 2.029% | 2006 |
| Xidian University | Xidian Univ | 61 | 1.925% | 2009 |
| University of Michigan | Univ Michigan | 58 | 1.830% | 2002 |
| Harvard University | Harvard Univ | 46 | 1.452% | 2003 |
| University of Illinois | Univ Illinois | 34 | 1.073% | 2006 |

As shown below in Figure 15, "the University London Imperial College of Science, Technology and Medicine" is the most active institution of the domain with strongest citation burst of 6.0668. The citation burst of this institution lasts for 5 years from 2010 to 2014. Whereas "the National Aeronautics and Space Administration" has longest citation burst, which has lasted for 15 years from 1994 to 2008. It affirms that NASA has been associated with a surge of citations.

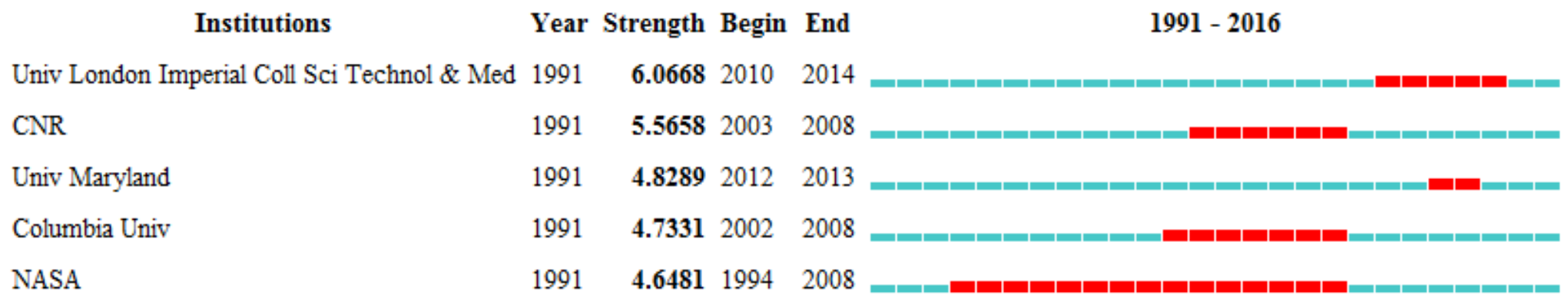

**Figure 15.** History of the burstness of the countries in the domain of "network community detection" in the timespan of 1991 to 2016. The "University of London Imperial College Science, Technology & Medicine" has strongest citation burst of 6.06668. The NASA has longest citation burst, which lasts from 1994 to 2008, whereas the "University of Maryland" has shortest citation burst.

After an overview of the visualisation of the collaborative institution network, next, we will have an overview of category co-occurrence network.

### 4.7. Category Co-Occurrence Network Analysis

This section presents the analysis of the co-occurrence of categories to identify articles associated with different categories. Figure 16 demonstrates temporal visualisation of the key categories in the bibliographic literature of community detection network.

The merged network of subject categories is comprised of 173 categories and 595 links. We have selected top 50 nodes for the slice length of one-year in the timespan of 1991 – 2016. In Figure 16, it can be easily seen that "Computer Science" is the highly-cited category and "Engineering" is the most central category of the domain. A detailed analysis is given underneath in the tabular form.

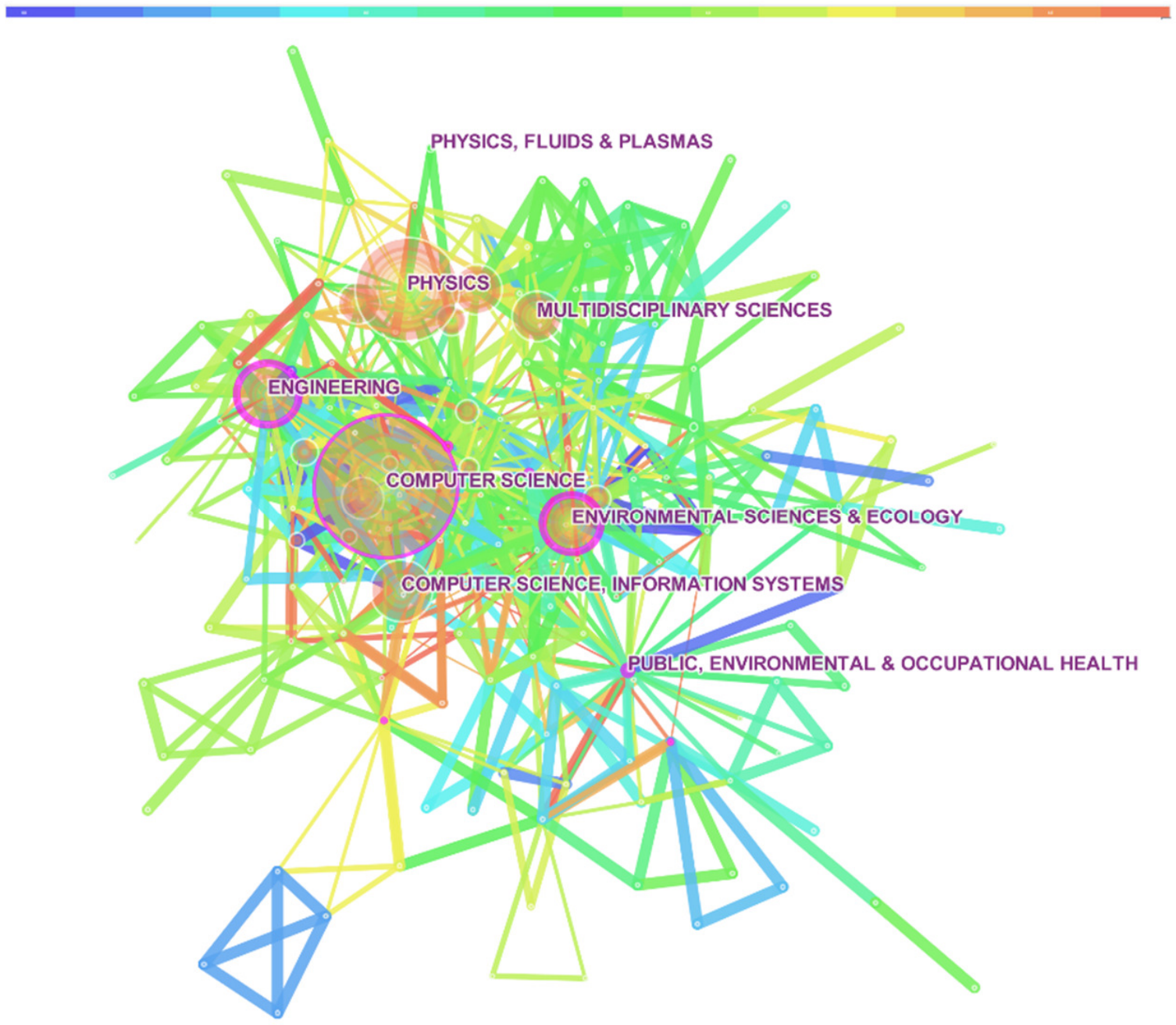

**Figure 16.** The visualisation of a network of categories in "network community detection" bibliographic data for the years 1991–2016. The merged network contains 173 nodes and 595 links. The concentric tree rings represent the temporal pattern of the categories in corresponding years. The purple circle around the node represents the betweenness centrality score, whereas red highlighted nodes represent the strength of citation burst. The colours of the links correspond to the particular time period.

Table 23 includes top five central categories of the domain. The category of "Engineering" is the most central category of the domain with the centrality score 0.37. Following it closely is the category of "Public, Environmental & Occupational Health" with the centrality score 0.35. Next is "Environmental Sciences" with a centrality score of 0.27. Next is the "Psychiatry" with a centrality score of 0.16. Finally, we have the category "Mathematics, Interdisciplinary Applications" with a centrality score of 0.14.

For comparative analysis, we have also performed analysis based on the frequency of the citations of the publications. The result of this analysis is demonstrated in Table 23.

**Table 23.** The top five categories based on the betweenness centrality in "network community detection" literature for the years 1991-2016. The "Engineering" is the most central category of the domain, whereas "mathematics, interdisciplinary applications" is the least central category of the domain.

| Category | Centrality | Year |
|---|---|---|
| ENGINEERING | 0.37 | 1991 |
| PUBLIC, ENVIRONMENTAL & OCCUPATIONAL HEALTH | 0.35 | 1992 |
| ENVIRONMENTAL SCIENCES | 0.27 | 1995 |
| PSYCHIATRY | 0.16 | 1996 |
| MATHEMATICS, INTERDICSIPLINARY APPLICATIONS | 0.14 | 2004 |

Table 24 presents top five highly occurred categories of the domain. The category of "Computer Science" with 724 occurrences leads over other categories of the domain, which is 22.853% of total records in the dataset. Next is the category of "Physics" with 530 occurrences, which is 16.729%. Then we have "Engineering" with 344 occurrences, which is 10.858%. It is closely followed by "Computer Science, Information Systems" with occurrences of 337, which is 10.637%. Subsequently, we have "Environmental Sciences & Ecology" with 317 occurrences, which is 10.006%.

**Table 24.** Top five categories based on frequency in "network community detection" bibliographic data for the years 1991 to 2016. The "Computer Science" is the highly-cited category of the domain. Whereas the "Environmental Sciences & Ecology" is least cited category of the domain.

| Category | Frequency | % of Records | Year |
|---|---|---|---|
| COMPUTER SCIENCE | 724 | 22.853% | 1991 |
| PHYSICS | 530 | 16.729% | 2001 |
| ENGINEERING | 344 | 10.858% | 1991 |
| COMPUTER SCIENCE, INFORMATION SYSTEMS | 337 | 10.637% | 1991 |
| ENVIRONMENTAL SCIENCES & ECOLOGY | 317 | 10.006% | 1991 |

Figure 17 demonstrates the history of the citation burst of the top five subject categories in the "network community detection" dataset. Burstness identify the subject categories which are active in the relevant research area. It also demonstrated the duration in which burst took place.

As shown in Figure 17, "PUBLIC, ENVIRONMENTAL & OCCUPATIONAL HEALTH" is the most active category of the domain with strongest citation burst of 13.9655. The citation burst of this category lasts for 13 years from 1991 to 2004. Two categories "Astronomy & Astrophysics" and "Medicine, General & Internal" have longest citation burst, which lasts for 17 years from 1993 to 2010. It provides evidence that these two categories have attracted a huge degree of attention from the research community of the domain.

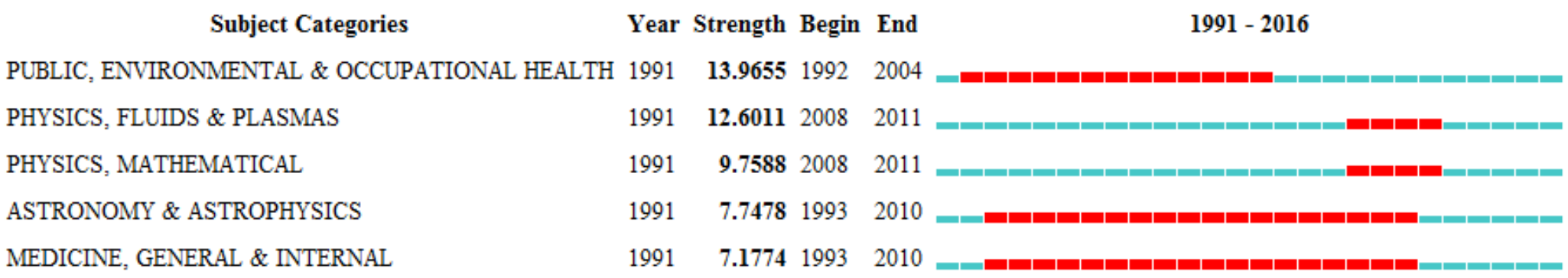

**Figure 17.** The citation history of the top five subject categories in the "network community detection" data. The "Public, Environmental & Occupational Health" category has highest citation burst. The categories of "Physics, Fluids & Plasmas" and "Physics, Mathematical" have shortest citation bursts. Whereas the categories of "Astronomy & Astrophysics" and "Medicine, General & Internal" have longest citation bursts.

After visualisation of co-citation network of authors, journals, and documents, the collaborative network of countries and institutions, and co-occurrence network of key categories, in the end, we are summarising the results.

## 5. Summary of the Results

In the current article, we have used CiteSpace for the comprehensive visual analysis of all pertinent peer-reviewed papers retrieved from WoS, devoted to network community detection over last 25 years. This section demonstrates an overview of the significant results obtained from scientometric analysis in this study.

Firstly, we acquire valuable information by identifying clusters of key authors, we found the "modularity" (cluster #0) is the largest cluster, which contains 52 vertices which are 3.6775% of total vertices in the network. The article of Zhang Y is the key pivot point, which connects "Potts Model; Time; Resolution" (cluster #3) and "non-overlapping community" (cluster #2). The article of Wang Y ties in diverse areas together. It joins "Modularity; Complex Network" (cluster #0), "Complex network; Overlapping Community Detection" (cluster #1), "non-overlapping community" (cluster #2), and "Excitable Systems; Complex Network" (cluster #31).

Successive analysis substantiated that there is conducted diversity in co-authors, co-cited authors, co-cited journals, co-cited documents, collaborative countries, collaborative institutions, and co-occurred subject categories.

In the author co-citation analysis, we observed that "Mark J. Newman" is the landmark node of the domain. We also observed that "Albert-László Barabási," "Pietro Perona," "Stephen Grossberg," and “Richard O. Duda” are the pivot nodes in the network. We also observed that “Réka Albert” has strongest citation burst.

In the co-citation analysis of journals, we identified that the “Reviews of Modern Physics” has the strongest citation burst. We also identified that “National Academy of Sciences of USA (PNAS)” is the most productive node with a high frequency of cited publications and “Science” is the most influential journal of the domain. We also observed that the cluster #1 (the second largest cluster), labelled as “Complex Network; Modularity; Community Structure” is the most influential, productive, and active area of the research. It contains most of the highly cited, highly central, and active journals of the domain.

In terms of the analysis of the document co-citation network, we observed that article by “Andrea Lancichinetti (2008)” is the most central document of the domain. We also observed that most cited article in the network is by “Forunato (2010).” We also found that cluster #0 is the major and active area of the research. Among most of the documents with the highest citation burst, the top five highly cited documents also belong to this cluster. Okasonen J (2001) is the key turning point which exclusively joins cluster #6 the “Biodiversity Hotspot” and cluster #11 the “Biological control agent.”

In the analysis of collaborative countries, top 30 countries per one-year time slice were selected from the timeframe $1991 - 2016$. We observed that the US has the highest frequency as well as the highest betweenness centrality, which indicates the origin of key publications in the domain. Scotland has the strongest citation burst, which affirms that the publications originating in the domain from Scotland have attracted a high degree of attention from the scientific community.

From the visualisation of collaborative institutions, we noted that “The Massachusetts Institute of Technology” has the highest betweenness centrality score the timeframe of 1991 to 2016. Whereas the “Chinese Academy of Sciences” has a top ranking with the largest frequency of publications among all other institutes. The “University of London Imperial College of Science and Technology and Medicine” has strongest citation burst.

Finally, in the analysis of co-occurrence of categories, we identified that the category “Engineering” leads overall categories in the domain with centrality value 0.37. Whereas with a frequency of 724, “Computer Science” leads the rest of the categories. We also identified that the category “Public, Environmental & Occupational Health.”

## 6. Conclusion

In this paper, CiteSpace is employed to trace advances in the field of “network community detection.” To this aim, we carried out a comprehensive visual scientometric analysis to assess research productivity and identify emerging trends. We have covered all relevant Journal articles indexed inS Thomson Reuters during the timespan of $1991 - 2016$. Our research is based on real data from the Web of Science databases. This permits us to comprehend all publications in the domain of “network community detection”. Our analysis has revealed many remarkable results. The “network community detection” has received the interest of its research community from the era of 1991, which accelerated after Newman’s article published in 2004. Santo Fortunato is the most highly cited author in the literature of community detection, whereas Réka Albert is the author who has rapidly grown the number of publications during the course of study. The “PNAS” is the most productive source journal; it contributed 52.935% publications during the period of study. The “United States” is the most productive and influential country, it has contributed the largest number of publications and has the highest centrality score. Most of the contributions in the domain came from “Chinese Academy of Sciences,” whereas the “University of London Imperial College of Science and Technology” remained specifically active in the research. The “Computer Science” leads the rest of the categories in the field.

Besides the conclusions, we believe that information and references from our analysis will provide a broader picture of the domain to the researchers. A significant dimension of future work is to achieve detailed insight through visual analysis of the subfields of the domain. Furthermore, our aim is to validate the findings our analysis by comparing with Pajek and by using other databases such as Scopus.